\documentclass{aa}  

\usepackage{natbib}
\bibpunct{(}{)}{;}{a}{}{,} % to follow the A&A style

\usepackage{txfonts}

\usepackage[english]{babel}
\usepackage{soul}
\usepackage{amsmath,amssymb}
\usepackage{graphicx}
\usepackage{subfigure}
\usepackage{xcolor}
\usepackage[colorlinks=True,allcolors=blue]{hyperref}
\usepackage[switch]{lineno} 
\begin{document} 

   \title{Erratum: Towards multi-messenger observations of core-collapse supernovae harbouring choked jets}
   \authorrunning{A. Zegarelli et al.}
   
\author{A. Zegarelli\thanks{E-mail: angela.zegarelli@astro.ruhr-uni-bochum.de}\inst{1}
\and
D. Guetta\inst{2}
\and
S. Celli\inst{3,4}
\and
S. Gagliardini\inst{2,3}
\and
I. Di Palma\inst{3,4}
\and
I. Bartos\inst{5}
}
\institute{
Ruhr University Bochum, Faculty of Physics and Astronomy, Astronomical Institute (AIRUB), Universitätsstraße 150, 44801 Bochum, Germany
\and
Department of Physics, Ariel University, Ariel, Israel
\and
Istituto Nazionale di Fisica Nucleare, Sezione di Roma, P. le Aldo Moro 2, I-00185 Rome, Italy
\and
Dipartimento di Fisica dell'Universit\`a La Sapienza, P. le Aldo Moro 2, I-00185 Rome, Italy
\and
Department of Physics, University of Florida, Gainesville, FL 32611-8440, USA
}
\abstract
{Over the past decade, choked jets have attracted particular attention as potential sources of high-energy cosmic neutrinos. It is challenging to test this hypothesis because of the missing gamma-ray counterpart. An identification of other electromagnetic signatures is therefore crucial. Extended H envelopes surrounding collapsing massive stars might choke launched jets. In addition, the same progenitors are expected to produce a shock-breakout signal in the ultraviolet (UV) and optical that lasts several days. Early UV radiation in particular carries important information about the presence and nature of choked jets.}
{While UV observations of core-collapse supernovae have so far been limited, the full potential of observations in this spectral band will soon be transformed by the ULTRASAT satellite mission with its unprecedented field of view.
We investigated the detection prospects of choked jet progenitors by ULTRASAT in relation to their visibility in the optical band by the currently operating telescope ZTF. In addition, as choked jets can produce neutrinos via hadronic and photohadronic interactions in choked jets, we also investigated how neutrino observations by existing Cherenkov high-energy neutrino telescopes (e.g. IceCube and KM3NeT) can be used in association with electromagnetic signals from shock-breakout events.}
{By considering fiducial parameters of the source population and instrument performances, we estimated the maximum redshift up to which ULTRASAT and ZTF are able to detect ultraviolet and optical signals from these explosions, respectively. Furthermore, we discuss coordinated multi-messenger observations using ULTRASAT, ZTF, and high-energy neutrino telescopes.}
{We find that ULTRASAT will double the volume of the sky that is currently visible by ZTF for the same emitting sources. This will enlarge the sample of observed SNe II by $\sim50\%$. For optimised multi-messenger detections, the delay between neutrinos produced at the shock breakout (SBO; during the jet propagation inside the stellar envelope) and ULTRASAT observations should be $\sim$4(5) days, with subsequent follow-up by instruments such as ZTF about one week later. We estimate that at most $\sim$20\% of the CCSNe from RSG stars that are detectable in the UV with ULTRASAT host a choked jet and release TeV neutrinos. Electromagnetic and neutrino detections, if accompanied by additional photometric and spectroscopic follow-up with compelling evidence of a relativistic jet launched by the central engine of the source, would suggest that CCSNe harbouring choked jets are the main contributors to the diffuse astrophysical high-energy neutrino flux.}
{}
%\abstract
% 5 {} token are mandatory

\maketitle
%
%-------------------------------------------------------------------
\section{Introduction}
\label{sec:introduction}
At the end of its life, a massive star (initial mass $\gtrsim8$ solar masses) will typically undergo core collapse, resulting in the formation of a compact object, a black hole or a neutron star. Accretion of some stellar matter by this compact object can drive an energetic relativistic outflow, or jet. The core collapse of massive stars that are stripped of their outer layers of hydrogen (H) and helium (He) by strong winds produces Type Ib/c supernovae (SNe). When these outer layers are not lost, core collapse results in a Type II SN.

For Wolf-Rayet (WR) stars, that is, very massive stars with strong winds that blow most of the stellar H envelope away, the relativistic jet driven by the central compact object can easily break through the remaining stellar envelope and produce $\gamma$-ray emission, which is usually observed as a burst lasting from a few to a thousand seconds. These phenomena are called long gamma-ray bursts (LGRBs). LGRBs are typically observed together with Type Ib/c SNe. In particular, LGRBs at low redshifts have been spectroscopically associated with broad-lined type Ic (Ic-BL) SNe (e.g. \citealt{cano2017}), which constitute a rare subclass of Ib/c SNe exhibiting very broad and blended spectral features due to high ejecta speeds (e.g. \citealt{mazzali}). On the other hand, powerful LGRBs are not associated with Type II SNe (hereafter, SNe II). This suggests that jets are unable to burrow through and escape the outer layers of SN II progenitor stars (see e.g. \citealt{2006ARA&A..44..507W} for a review about the connection between GRB and SN). 

H-rich SNe II come from the explosion of extended massive stars. Because of the wide variety of observational properties in their light curves and spectra, these can be further divided into several categories (see \citealt{lin_universe2023} for a recent review about the diversity of SNe II). Type II-P and II-L are defined according to whether their light curve shows a plateau or decays linearly. In Type IIb SNe, H features are present in the early days, but diminish over time as helium lines gain in strength. Finally, the spectra of Type IIn SNe exhibit narrow (a few hundred km/s) H emission lines that are caused by the interaction of the supernova shock with the circumstellar medium (CSM), namely with dust and gas around the star.
Since for most of them only a few progenitor detections are available, we lack enough observational evidence to draw conclusions about the whole classes. In addition, there is still uncertainty about their progenitors, except for Type IIP SNe.
These are the most common of all core-collapse SNe (CCSNe). They account for nearly 60\% of the total number of SNe II, and detections in pre-SN images have allowed the community to firmly associate them with red supergiants (RSGs).
Among the possible progenitors of CCSNe, it is also worth considering blue supergiants (BSGs) \citep{bsg_parameters}.  Rare under-luminous SNe II, such as SN 1987A \citep{sn1987a}, may arise from them. BSGs may also be the progenitors of so-called ultra-long GRBs (ULGRBs), which are characterised by a $\gamma$-ray emission lasting $\gtrsim10^4$ seconds \citep{2018ApJ...859...48P}, that is thought to be generated by fallback accretion from the extended stellar envelope \citep{2012MNRAS.419L...1Q,2013ApJ...767L..36W,2013ApJ...778...67N}.

The large radial extension of both RSGs and BSGs constitutes a challenge for the emergence of the jet \citep{2018ApJ...859...48P}, as the preservation of their large stellar envelope (a few dozen solar radii, or $R_{\odot}$) might lead to choked jets when the central engine activity stops early enough before the jet reaches the outer edge of the star. If this is so, $\gamma$-ray emission would not emerge, as the remaining envelope is opaque to $\gamma$ rays. Simulations show that stellar envelopes of non-rotating RSG stars are expected to be even larger (between $\sim 200~R_{\odot}$ and $\sim 1500~R_{\odot}$; e.g. \citealt{rsg_parameters}). This causes a very long accretion timescale ($\sim 10^5-10^6$ s) and a low accretion rate of material around the central engine (e.g. \citealt{2018ApJ...859...48P}).
This may drive a weak jet that is unable to burrow through the thick stellar envelope of the RSG. 
The jet might dissipate all its energy before reaching the edge of stars without producing any GRB. Because of this, it is reasonable to consider RSGs and BSGs as candidate progenitors for failed GRBs.
It is worth mentioning that CSM effects have also been suggested, analogously to the choked-jet hypothesis, to explain low-luminous GRBs (\textit{ll}GRBs), which are characterised by an isotropic $\gamma$-ray luminosity of $E_{\gamma,\rm iso} \lesssim 10^{49}~\rm erg$ \citep{Senno2016}. \textit{ll}GRBs may arise from jets that cannot break out of the star, thus failing to power the prompt emission as in LGRBs, but are still capable of producing a detectable signal (e.g. \citealt{llGRB_TypeIbc}). The few associations existing between SNe and \textit{ll}GRBs so far connect the latter to Ib/c SNe. These GRBs would then be surrounded by an envelope that is larger than the typical size of WR stars, but not as extended as RSGs and BSGs. A representative case of this class of sources is the nearby \textit{ll}GRB 060218 ($z=0.033$, $L_{\gamma} \sim 10^{47}~\mathrm{erg~s^{-1}}$) associated with \textit{SN 2006aj} \citep{Nakar2015}. 

Relativistic jets launched by a core collapse are also expected to produce a cocoon due to the jet depositing energy in the stellar envelope \citep{1999ApJ...524..262M}. This is expected to transport an amount of energy comparable to that of the prompt GRB emission. As the GRB jet carves its way through the stellar envelope, it dissipates its energy in a double shock (forward-reverse) structure that forms at its head \citep{2003MNRAS.345..575M,2005ApJ...629..903L,2011ApJ...740..100B}. The hot head material spills sideways, forming a cocoon that engulfs and collimates the jet.
The first observational evidence for a cocoon came from the long GRB 171205A \citep{2019Natur.565..324I}, which occurred at the unusually small distance of 163\,Mpc. Detailed multi-epoch spectroscopic observation of the associated \textit{SN 2017iuk} allowed the identification of the cocoon emission as an additional blackbody component in the light curve about one day after the GRB.
Spectroscopic observations of the early-time SN spectrum revealed sub-relativistic material ($v\sim c/3$) rich in Fe and Ni, indicating that the observed component comes from deep inside the star. The implications of the presence of the cocoon, as well as more details about the theoretical framework related to it, were discussed in \cite{Nakar2017}.

When the jet crosses a large fraction of the stellar envelope before it stalls, the cocoon can be energetic enough even for failed GRBs to break away from the star by itself and produce an observable electromagnetic signature: A bright X-ray/UV flash that lasts from few seconds to a fraction of an hour, followed by a long-lasting (day timescale) UV and optical emission related to the expanding cocoon envelope. In any scenario, however, the electromagnetic signal requires that the shock reaches the stellar surface. We refer to \cite{2017hsn..book..967W} for details on shock breakout (SBO) theory. 

Prior to the electromagnetic emission, high-energy neutrinos may also be produced by the hidden jet: Hadronic acceleration and subsequent interaction with the intense radiation field produced in the SN explosion lead to neutrinos at TeV-PeV energy scale that are immediately released \citep{2001PhRvL..87q1102M}. In contrast, $\gamma$ rays emitted in the jet are likely absorbed in pair-production processes because of the intense radiation fields intrinsic to the source. In this framework, high-energy neutrinos arise from a process that is different from the process that causes thermal MeV neutrino production at the core collapse of massive stars for instance via electron-captor and pair annihilation on a timescale of $\mathcal{O}(10)$~s from the core bounce (see e.g. \citealt{neutrinoSN_review} for a review of neutrino production from supernovae). Neutrinos from choked jets are characterised by higher energies, are described by a non-thermal spectrum, and are emitted at later times. The TeV-PeV neutrino signal from choked jets was recently considered by several authors \citep{2012PhRvD..86h3007B,MuraseIoka2013,He2018,senno2018,stacking_choked,fasano,Chang2022}. In the presence of a sufficiently dense medium, $pp$ collisions can also occur, generating further neutrinos via meson and muon decay.

In this work, we focus on SN types coming from stellar envelopes without significant CSM. In particular, we consider BSG and RSG progenitors, assuming that both of them embed choked relativistic jets in their stellar envelopes (e.g. \citealt{1999ApJ...524..262M,2001PhRvL..87q1102M,He2018,fasano,2023A&A...679C...1R}). We investigate whether the corresponding SBO transients can be followed up with current and future facilities via a multi-messenger approach that allows the community to combine neutrino, optical, and UV data. Defining a proper follow-up strategy between these messengers could be the key for enabling the breakthrough discovery of the astrophysical sources that cause some of the observed cosmic neutrinos. The inconsistency between the cosmic neutrino flux and the isotropic $\gamma$-ray background observed by the \textit{Fermi} satellite indeed indicate hidden sources as plausible contributors. If the entirety of neutrino sources were transparent to energetic photons, the expected $\gamma$-ray flux would exceed the diffuse extragalactic flux observed by the \textit{Fermi} Large Area Telescope (LAT), requiring at least some neutrino sources to be opaque to $\gamma$ rays \citep{Ackermann2015,fermi_inconsistence}. In addition, improved observations and an analysis of choked jets can shed new light on the dynamics and evolutionary path of exploding massive stars. If all SNe are accompanied by SBO, detecting SBO transients and their signatures might be used to reveal the properties of SN progenitors. UV observations from CCSNe, for example, are expected to better constrain physical stellar properties such as progenitor radius, surface composition, and explosion energy per unit mass \citep{Ganot2016}. 

The paper is organised as follows. In Section~\ref{sec:emissions_sbo}, we review the main features of the model under investigation, including RSGs and BSGs. In Section~\ref{sec:detection_sbo}, we describe the method we developed for assessing the detectability of UV signals from their choked jets and present the results we obtained. In Section~\ref{sec:mm_followups}, we discuss the implementation of a multi-messenger framework tailored to maximising the observation prospects. Finally, we conclude in Section~\ref{sec:conclusions}.
Throughout, we adopt the standard flat $\Lambda$ cold dark matter ($\rm \Lambda CDM$) cosmology with $\Omega_m=0.3$, $\Omega_{\Lambda}=0.7$, and $H_0=70~\rm km~s^{-1}~Mpc^{-1}$.

\section{UV and optical emission in shock breakout flares from hydrogen-dominated stellar envelopes}
\label{sec:emissions_sbo}
The first electromagnetic emission escaping an exploding star emerges as a fast shock-breakout flare, with a spectrum that peaks in the UV and X-ray bands. After the breakout, the stellar envelope expands and cools. As the photosphere penetrates the outer shells of the envelope, the adiabatically cooled radiation stored within the envelope escapes and leads to early UV and optical emission over a timescale of days. Several analytic and numerical solutions have been developed that are capable to model the emission at early times, that is, $<3$ hours post explosion (e.g. \citealt{ns10,swk13}) and at later times (days after the explosion) \citep{rw11}.

The RSG, BSG, and WR stars differ greatly in their UV peaks. Early UV observations are therefore a strong discriminator among progenitor classes (see Figure~1 in \citealt{Ganot2016}). We focus on choked jets here and therefore limit the discussion to RSGs and BSGs, which are both characterised by H-dominated envelopes. WR stars with He- or carbon- or oxygen-dominated envelopes instead lose their H envelope, allowing the jet to break out (see Section~\ref{sec:introduction}). 
We modelled UV and optical cooling after the explosion  following \cite{waxman2007}, which we describe in the following. %the next Section.

%\subsection{A simplified model for H-dominated envelopes}
%\label{sec:simple_model}
In H-dominated stellar envelopes without a significant CSM and characterised by an adiabatic flow below the photosphere, the radius $r_{\mathrm{ph}}$ and the effective temperature $T_{\mathrm{ph}}$ of the photosphere penetrating the outer shells of the envelope are given by \citep{rw11}
\begin{equation}
\label{eq:phradius}
r_{\mathrm{ph}}(t)=\left\{\begin{matrix}
3.3\times 10^{14}~f_{\rho}^{-0.062}\frac{E_{51}^{0.41}k^{0.093}_{0.34}}{(M/M_{\odot})^{0.31}}t_5^{0.81}~\mathrm{cm} & (n=3/2),\\ 
\\
3.3\times 10^{14}~f_{\rho}^{-0.036}\frac{E_{51}^{0.39}k^{0.11}_{0.34}}{(M/M_{\odot})^{0.28}}t_5^{0.78}~\mathrm{cm} & (n=3),
\end{matrix}\right.
\end{equation}
and
\begin{equation}
\label{eq:tph}
T_{\mathrm{ph}}(t)=\left\{\begin{matrix}
1.6~f_{\rho}^{-0.037}\frac{E_{51}^{0.027}R _{*,13}^{1/4}}{(M/M_{\odot})^{0.054}k_{0.34}^{0.28}}t_5^{-0.45}~\mathrm{eV} & (n=3/2),\\ 
\\
1.6~f_{\rho}^{-0.022}\frac{E_{51}^{0.016}R _{*,13}^{1/4}}{(M/M_{\odot})^{0.033}k_{0.34}^{0.27}}t_5^{-0.47}~\mathrm{eV} & (n=3),
\end{matrix}\right.
\end{equation}
respectively, under the assumption that the photon diffusion is negligible. The effect of photon diffusion was estimated by \cite{rw11}, who derived the radius above which diffusion affects the flow significantly and estimated the effect on the temperature and luminosity of the approximation considered in \cite{waxman2007}. This effect would alter our model results only by $\sim 1\%-5\%$, and we therefore neglect it below.
In previous equations, $k=0.34k_{0.34}$~cm$^{2}$~g$^{-1}$ is the opacity of the stellar envelope,  $E=10^{51}E_{51}$ erg is the kinetic energy released at the SN explosion, $t=10^5 t_5$ s is the time from the SBO, and $f_{\rho}$ is a numerical factor of order unity that depends on the inner envelope structure and that is linearly related to the density profile of a polytropic envelope. This latter quantity scales linearly with the progenitor mass and can be described by a power law with index $n=3$ for radiative envelopes typical of BSGs, and $n=3/2$ for efficiently convective envelopes typical of RSGs. 
We considered a stationary and uniform opacity that applies for example to Thomson-scattering dominated opacity with constant ionisation, as in fully ionised H-dominated envelopes where $k=0.34$~cm$^{2}$~g$^{-1}$.
The parameters in Equation~\eqref{eq:phradius} and Equation~\eqref{eq:tph} have a strict relation with the expected specific intensity in UV and optical wavelengths. This latter is given by a blackbody (BB) radiation as modified by extinction, namely the absorption and scattering of electromagnetic radiation by dust and gas intervening along the line of sight, yielding
\begin{equation}
\label{eq:flambda}
    f_{\lambda}(\lambda,t)=\left ( \frac{r_{\mathrm{ph}}}{D_L(z)} \right )^2 \sigma T_{\mathrm{ph}}^4 \frac{T_{\mathrm{col}}}{hc}g_{BB}(x)e^{-\tau_{\lambda}},
\end{equation}
where $D_L(z)$ is the luminosity distance of the source dependent on redshift $z$, $T_{\mathrm{col}}$ is the colour temperature, that is, the temperature at which a BB would emit radiation of the same colour as the given source, $\tau_{\lambda} \equiv \tau(\lambda)$ is the optical depth at a given wavelength $\lambda$, $\sigma$ is the Stefan–Boltzmann constant, $h$ is the Planck constant, $c$ is the speed of light in vacuum, and
\begin{equation}
    g_{BB}(x)=\frac{15}{\pi^4}\frac{x^5}{e^x-1}
\end{equation}
with $x=hc/\lambda T_{\mathrm{col}}$. Starting from Equation~\eqref{eq:flambda}, the light curve visible in the UV can be expressed as
\begin{equation}
    L_{UV}(t)=4\pi D_L^2(z)\int_{\lambda_{min}}^{\lambda_{max}}f_{\lambda}(\lambda,t)d\lambda.
\end{equation}
By using representative values for the photospheric radius and temperature of stellar progenitors (in Equations~\eqref{eq:phradius} and \eqref{eq:tph}, respectively), we can estimate the intensity of UV light emitted at a given time from the SBO, and we can hence characterise the corresponding light curves.

The simplified model from \cite{waxman2007} is valid in a Thomson-scattering dominated regime, namely at early times after the SBO. It was extended by \cite{rw11} to take into account that recombination takes place in stellar envelopes after about one day, when $T_{\mathrm{ph}}$ decreases down below $\sim 1$~eV. Moreover, the authors introduced a more realistic characterisation of the opacity (including its time dependence) as modified by recombination, thus obtaining a more accurate description of the early UV and optical emission produced in SBO phenomena. However, for H-dominated envelopes, the difference in the values of photospheric temperature obtained by the two methods is contained within 10\% for $T_{\mathrm{ph}}>1$ eV. We therefore decided to adopt the model of \cite{waxman2007} neglecting any complications related to recombination processes. 

\section{Detection of shock-breakout electromagnetic signals from CCSNe}
\label{sec:detection_sbo}
With the launch of the \textit{Neil Gehrels Swift} Observatory \citep{swift} in 2005, a large amount of data of CCSN light curves at peak emission was collected through its rapid slewing capabilities, space-based nature, and frequency coverage in X-ray and O/UV. Since then, the number of observed SBO events has dramatically increased, opening a new window for early SN observations, that is, immediately after the explosion of the massive progenitor (e.g. \citealt{2009ApJ...702..226M,2010AAS...21534203R,2018MNRAS.475.3959H,pritchard2014}). These observations were often associated with further detections made in the UV, visible, and near-infrared regions of the electromagnetic spectrum for example by the \textit{Hubble} Space Telescope (HST) \citep{HST_CCSNe} in the optical domain and by the Galaxy Evolution Explorer (GALEX) satellite in UV wavelengths \citep{Ganot2016,ganot2}.
Joint studies have allowed the astronomical community to improve the characterisation of
%progenitors of choked jets, namely RSG and BSG, as well as to recognize in
CCSNe light curves, including the signatures of SBOs. In this regard, GALEX was a pioneering project. After the very first observations of SBOs from  SNe II in 2008 (\textit{SNLS-04D2dc} and \textit{SNLS-06D1jd}, \cite{firstCCSNe_UV}) through the combination of its UV data with optical observations from other instruments, it further allowed the astronomical community to increase the sample of CCSNe detected in both optical and UV wavelength band, which might stem from SBO events (see e.g. \cite{ganot2}).

The growing field of multi-messenger astrophysics enables us to increase the detection prospects of choked jets (which are invisible to $\gamma$ rays) through the additional inclusion of neutrino observatories.
The association of a neutrino emission with optical/UV signals could be crucial for probing the contribution of choked jets embedded in CCSNe to the diffuse flux of high-energy astrophysical neutrinos observed for more than ten years to date \citep{icecube_diffuse1,icecube_diffuse2}. While $\gamma$ rays are commonly adopted for joint studies of astrophysical sources with neutrino data due to a potential common hadronic origin, for hidden sources, it instead appears to be crucial to establish an alternative strategy that profits of signatures at other wavelengths.
To date, optical and X-ray follow-up program searches for these transient sources have been performed so far as a result of alerts from neutrino telescopes operating in the past decade (e.g. \citealt{antares_followup,icecube_multiplets}), but these mostly concerned neutrino alerts transmitted in the case of events clustering in space and time (two or more muon neutrino candidates in directional coincidence and arriving within a predefined time window), rather than starting from an X-ray/optical transient.

Alternative electromagnetic counterparts to $\gamma$ rays might arise from the UV and optical cooling emission after SBO phenomena in CCSNe.
In this regard, it is worth noting that in contrast to optical wavelengths, the UV band is not well covered by currently operating instruments. This shortage will soon be mitigated by the launch of a new satellite sensitive to the near-UV (NUV) band, called Ultraviolet Transient Astronomy Satellite (ULTRASAT) \citep{ultrasat}, which is planned for 2026. ULTRASAT will carry a UV telescope with an unprecedented field of view (FoV) of 204 deg$^2$. For comparison, the celestial volume monitored by ULTRASAT will be larger by 300 times than that of GALEX. It will be able to detect emission in the near-UV band (230-290) nm, reaching a mean limiting magnitude of $m_{\mathrm{AB}}^{\rm lim,USAT}=22.5$ after 900 s of observation time at a statistical significance of $5\sigma$\footnote{In the central 170 deg$^2$ of the FoV and for a BB source of $T_{\mathrm{col}}=20000$~K.}. We quantify the improvement that the scientific community will gather when ULTRASAT will be operational by combing its results with those of other existing facilities, such as the Zwicky Transient Facility (ZTF) for the optical band \citep{ZTF_sciencepaper,ztf_observing_system}. The latter has been surveying the Northern sky  since June 2018 every two to three nights in the $\textit{g}$-band at (370-560)~nm and $\textit{r}$-band at (550-740)~nm filters, while the $\textit{i}$-band filter at (690-895)~nm is used for partnership observations only. The ZTF is characterised by an instantaneous FoV of $\sim47~\rm deg^2$ that represents the largest instantaneous field of view of any camera on a telescope of aperture greater than 0.5 m. We consider its $5\sigma$ median sensitivity over all filters ($\textit{g}$, $\textit{r}$, and $\textit{i}$ bands) and all lunar phases for 30 seconds exposure, that is, $m_{\rm AB}^{\rm lim,ZTF}=20.4$ \citep{ztf}. As at the moment of writing the future of the ZTF is still uncertain, that is, it is unclear whether it will be operational by the time of the ULTRASAT launch, the results obtained and discussed throughout the paper in terms of ZTF have to be read as related to ZTF-like instruments.
Our results are intended to provide guidance for the scientific community for future multi-messenger studies involving $\nu$s, UV, and optical signals. In Section~\ref{sec:rate} we therefore explain the method we adopted to estimate the detectability of shock-cooling UV and optical emission from choked-jet progenitors with ULTRASAT and ZTF, and in Section~\ref{sec:nevents} we report the  corresponding results.

\subsection{Shock-cooling emission from choked-jet progenitors: Detectability with ULTRASAT and ZTF}
\label{sec:rate}
Starting from a set of fiducial progenitor parameters for BSGs and RSGs, we estimated the feasibility of combined observations between ULTRASAT and ZTF in order to quantify the expected improvement in discovering signals from the shock-cooling emission of choked-jet SNe II. 
We proceeded with the comparison of available instrument responses and model expectations about the specific intensity emissions resulting from SBOs, as described in Equation~\eqref{eq:flambda}.
For RSGs, we adopted progenitor parameters that were directly constrained from SN II observations.
In particular, we started from the analysis in \cite{ganot2} of a sample of CCSN light curves characterised by clear NUV flares and optical detection in the $R$ band, from simultaneous GALEX and Palomar Transient Factory (PTF) \citep{ptf1,ptf2} measurements between 24 May 2012 and 28 July 2012. \cite{ganot2} fitted the SN sample with the same model as we considered here. By averaging the best-fit values of \cite{ganot2} analysis, we therefore obtained the following values, which we adopted as representative of the RSG population: $R_{*}=722~R_{\odot}$, $E=10^{51}~\rm erg$, $M_{\rm ej}=2.8~M_{\odot}$, and $f_{\rho}=1.455$. These parameters agree with theoretical and computational predictions (e.g. \citealt{rsg_parameters}). For BSGs, we in turn relied on modelling expectations, because these sources are still unconstrained by observations. Typical parameters for BSGs that we set as representative of this population of sources are $R_{*}=50~R_{\odot}$, $M_{\rm ej}=10~M_{\odot}$, $f_{\rho}=0.0465$, and $E=10^{51}~\rm erg$ (e.g. \citealt{bsg_parameters}). For completeness, we investigate in Appendix \ref{appendix:uncertainties} the impact of these parameters on the signals expected from the models.
\begin{figure}[t!]
    \resizebox{\hsize}{!}{\includegraphics{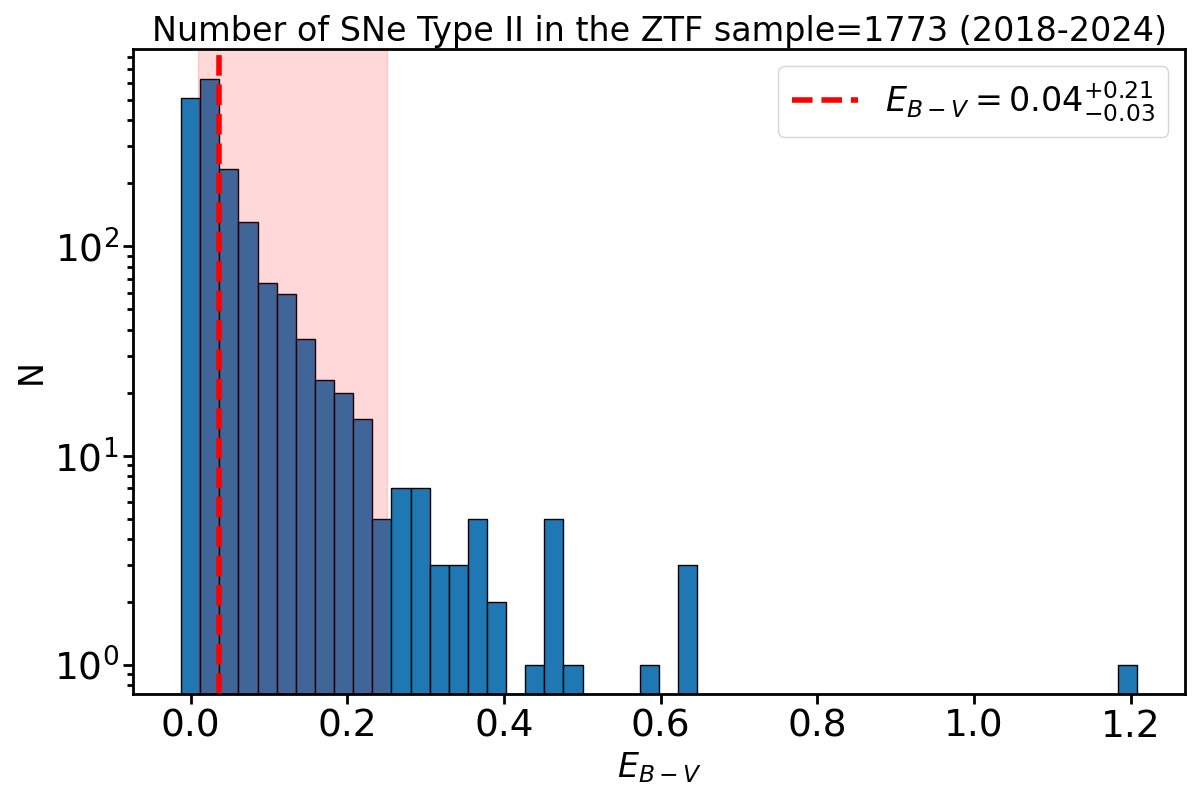}}
    \caption{Distribution of Galactic extinction values for galaxies hosting SNe II detected by ZTF (catalogue available online in \url{https://sites.astro.caltech.edu/ztf/bts}). The sample contains 1773 SNe II available in the ZTF Bright Transient Survey between June 2018 and the moment of writing (mid-February 2024). The dashed red line and the corresponding shaded region represent the median value of the distribution and the $2\sigma$ uncertainty band, obtained via the 2.28th and 97.72nd percentiles.}
    \label{fig:extinction}
\end{figure}
\begin{figure*}
\subfigure[\label{fig:a}]{\includegraphics[width=1\columnwidth]{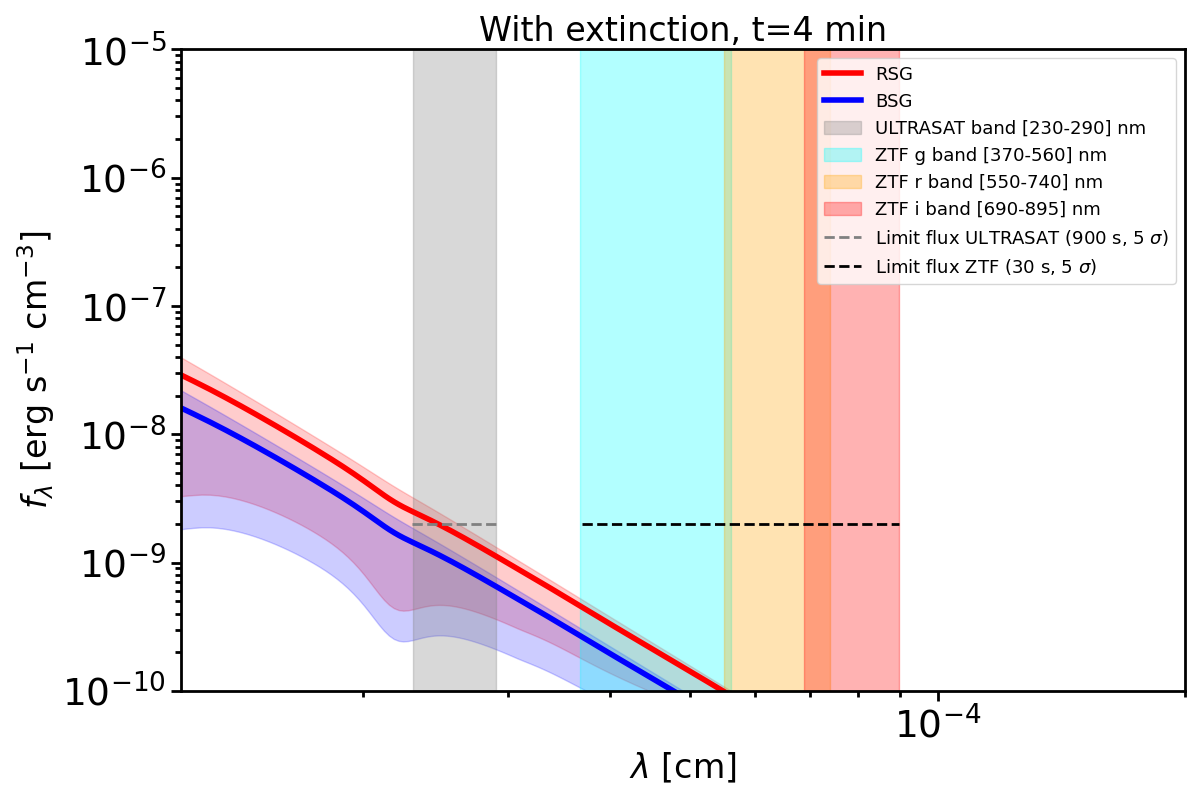}}
\subfigure[\label{fig:b}]{\includegraphics[width=1\columnwidth]{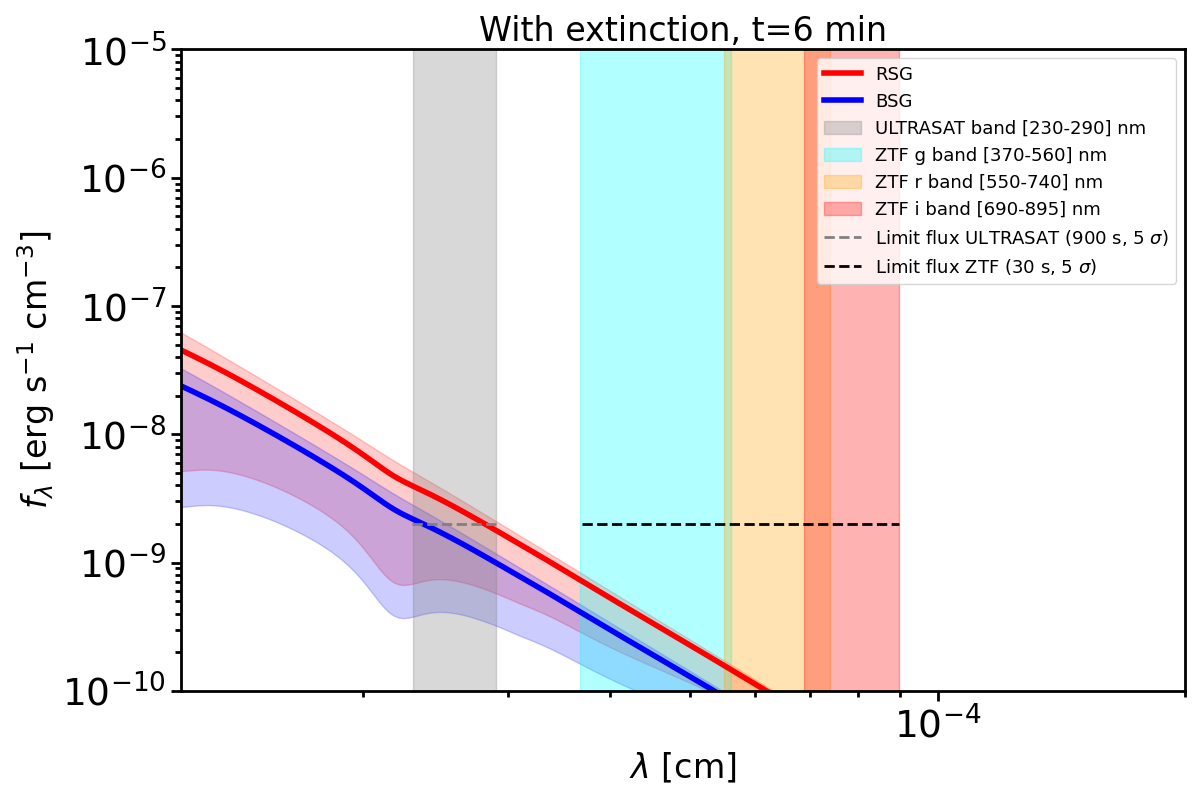}}\\
\subfigure[\label{fig:c}]{\includegraphics[width=1\columnwidth]{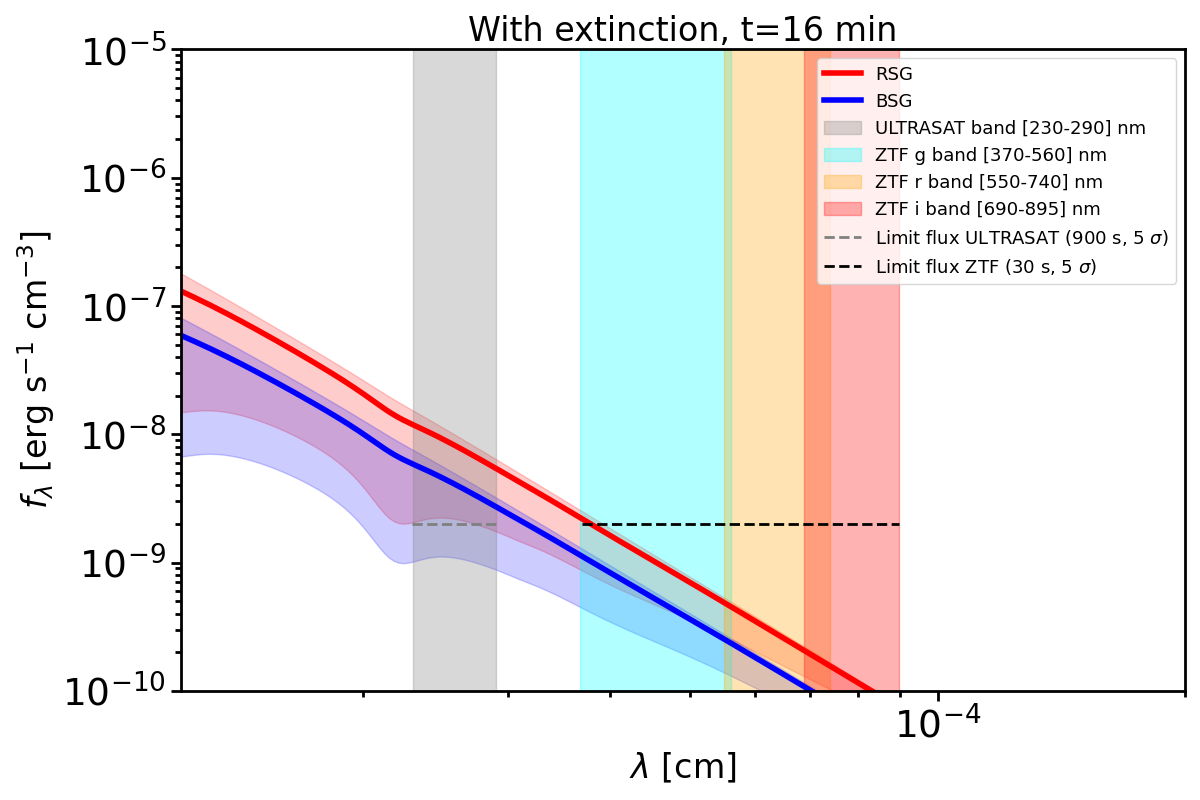}}
\subfigure[\label{fig:d}]{\includegraphics[width=1\columnwidth]{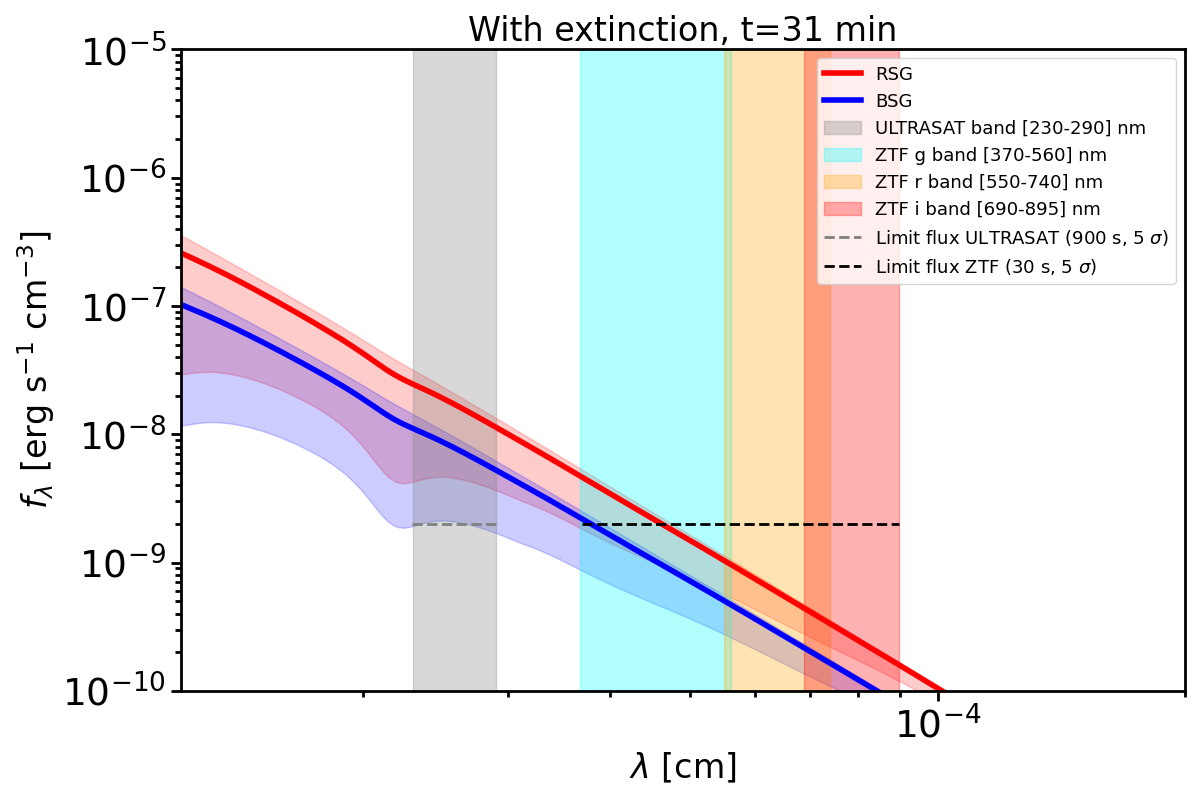}}\\
\subfigure[\label{fig:e}]{\includegraphics[width=1\columnwidth]{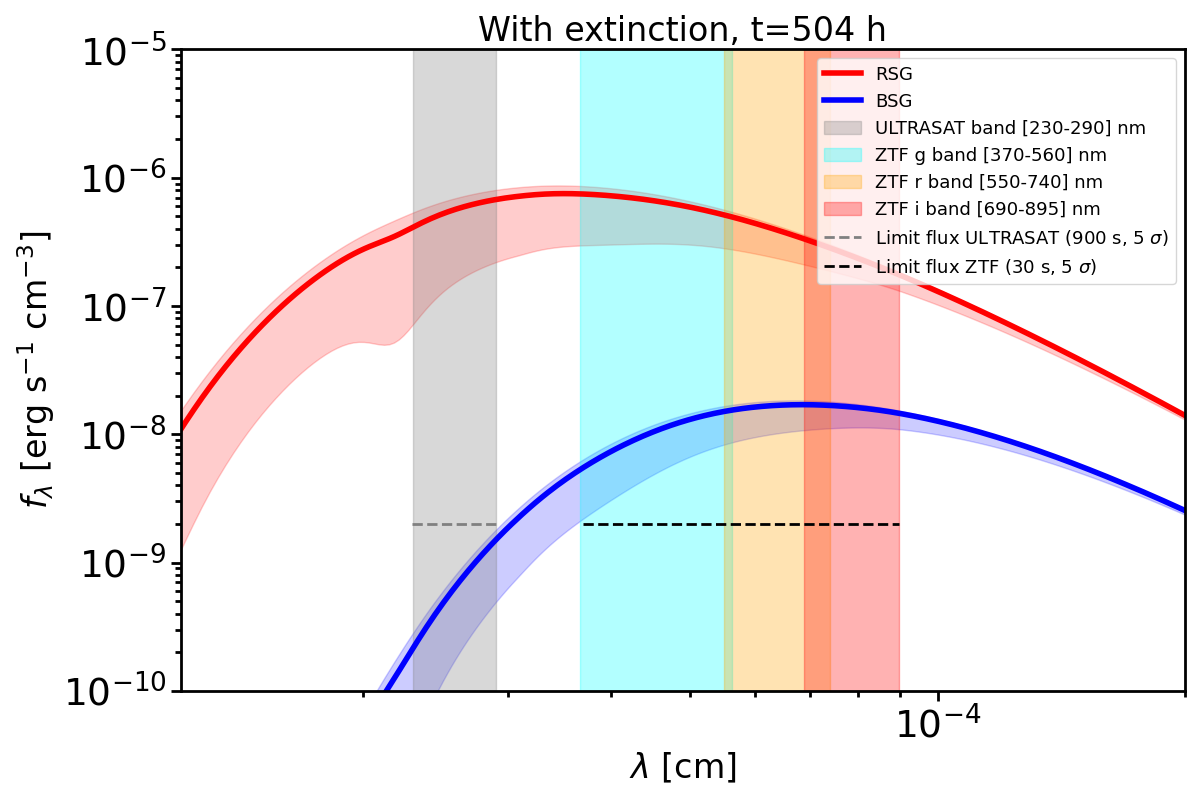}}
\subfigure[\label{fig:f}]{\includegraphics[width=1\columnwidth]{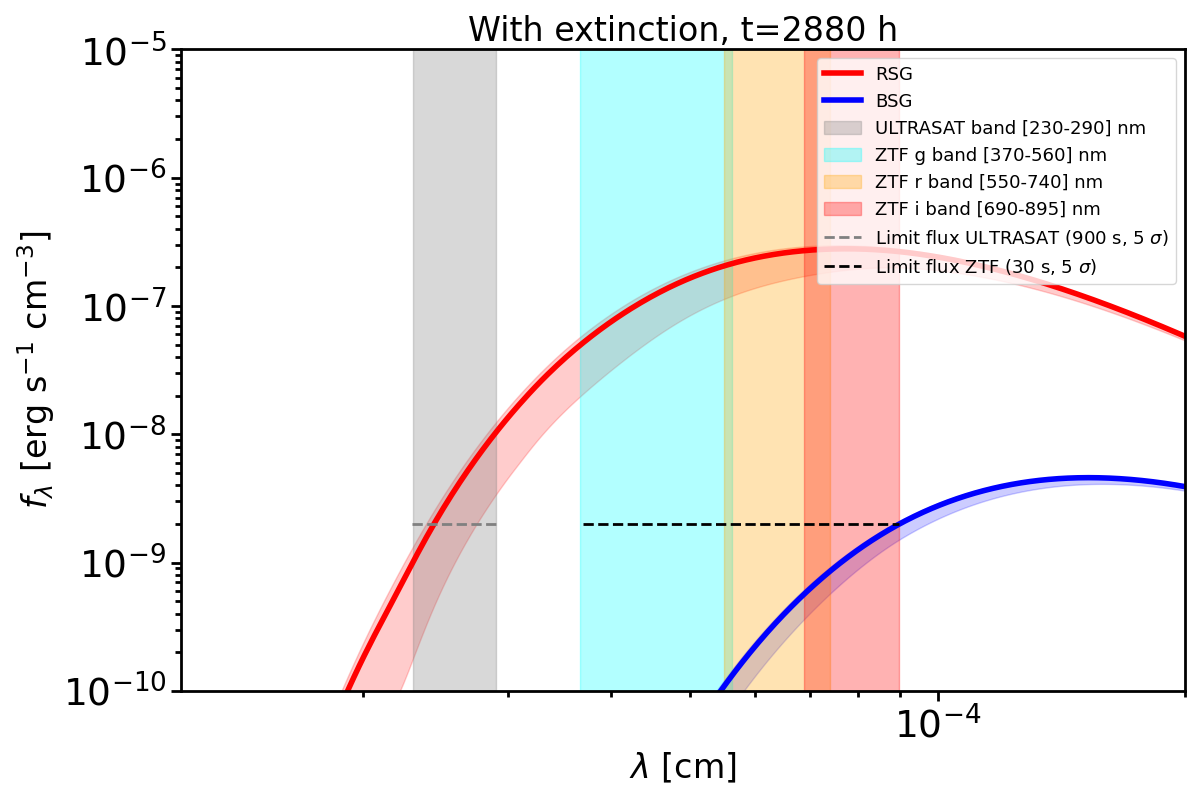}}\\
\caption{Extinction-corrected specific intensities, $f_{\lambda}$, observable at Earth from a RSG (in red) and BSG (in blue) located at $z=0.01$, as expected from the model in Eq.~(3). The fiducial parameters used in the computation for RSGs are $R_{*}=722~R_{\odot}$, $E=10^{51}~\rm erg$, $M_{\rm ej}=2.8~M_{\odot}$, and $f_{\rho}=1.455$, and for BSGs $R_{*}=50~R_{\odot}$, $M_{\rm ej}=10~M_{\odot}$, $f_{\rho}=0.0465$, and $E=10^{51}~\rm erg$. The solid lines show the expected emission with a median Galactic extinction $E_{B-V}=0.04$, and the corresponding shaded region the minimum and maximum extinction values, i.e. $E_{B-V}=0.01$ and $E_{B-V}=0.25$, respectively (see Fig.~1 in the original manuscript). The vertical shaded bands represent the wavelength range covered by each instrument: ULTRASAT (in grey) and ZTF with the $\textit{g}$ filter (in cyan), $\textit{r}$ filter (in yellow), and $\textit{i}$ filter (in red). The horizontal dashed grey and black lines define the $f_{\lambda}^{\rm lim}$ of the two instruments, obtained following Eq.~(7).
Panels (a) to (f) show the different evolution times, from $\sim$ 5 minutes to $\sim$ 4 months after the SBO event, highlighting for each stellar progenitor the time range when the emission was visible to ULTRASAT and ZTF. The temporal evolution from 1 minute to 1 year of $f_{\lambda}$ coming from RSGs and BSGs shown here is available as an online animation (supplementary material in the digital version of the paper).}
\label{fig:evolution_time_comparison_bsg_rsg}
\end{figure*}

With regard to the extinction term in Equation~\eqref{eq:flambda}, our analysis includes its dependence on wavelength in order to properly predict the observable UV and optical fluxes. In other words, we considered the variation in the total extinction value with wavelength $\lambda$, that is, $A_{\lambda}=1.086\tau_{\lambda}$, rather than using a constant value. This choice was motivated by the fact that the extinction is very pronounced at UV wavelengths, being characterised by a bump at $220$~nm, which is close to the ULTRASAT observational band. The total extinction in the visual $V$ band is defined as $A_V=R_V E_{\rm B-V}$, where $E_{\rm B-V}=E_{\rm B-V}^{\rm observed}-E_{\rm B-V}^{\rm intrinsic}$ represents the observed colour excess, and $R_V$ characterises the interstellar extinction from the near-infrared to the far-UV spectral region. We adopted the averaged dust extinction model of \cite{cardelli} for a diffuse interstellar medium in the Milky Way with $R_V=3.1$ from $\lambda_{\rm min}=0.1~\mu$m to $\lambda_{\rm max}=3.3~\mu$m.
To realistically quantify the impact of extinction in our calculations, we used public data available from the ZTF Bright Transient Survey (BTS)\footnote{The sample is updated daily; results are available in \url{https://sites.astro.caltech.edu/ztf/bts}.}. 
By selecting the SN Type II reported in the catalogue, we computed the median value of Galactic extinction characteristic of galaxies hosting the SNe II observed so far by ZTF. At the moment of writing, the sample contained 1773 sources classified as SN II for a median extinction in their host galaxies of $E_{B-V}=0.04^{+0.21}_{-0.03}$ (see Figure~\ref{fig:extinction}). Therefore, we consider the median $E_{B-V}=0.04$ as a benchmark value in the following evaluations and finally include an uncertainty at $2\sigma$  with $E_{B-V}=0.01$ and $E_{B-V}=0.25$ for the minimum and maximum extinction models presented in the following.
%\footnote{Extinction in dense molecular clouds in the Galaxy are characterised by larger values of $R_V$ higher (e.g., \citealt{molecolar_clouds_extintion_1,molecolar_clouds_extintion_2}); however, molecular clouds cover a small fraction of the sky.}).
%with extinction value $E_{\rm B-V}=0.04$ \textcolor{red}{referenza}. \textcolor{red}{For more details about extinction see Appendix XX.}

%\subsection{The wide-field UV space telescope ULTRASAT: properties}
%\label{sec:ultrasat}
After correcting for extinction, we investigated the possible detectability of the estimated cooling emission signals from SBOs by comparing them to the sensitivity of each instrument expressed as the limiting magnitude in the AB system $m_{\rm AB}^{\rm lim}$. This is related to the minimum observable flux spectral density via the relation
\begin{equation}
    \label{eq:mab_fnu}
    m_{\mathrm{AB}}=-2.5\log_{10}\left( \frac{f_{\nu}}{\rm [Jy]}\right)+8.90,
\end{equation}
which in turn can be converted into the corresponding spectral flux density per unit wavelength $f_{\lambda}$ as
\begin{equation}
\label{eq:fnu_flambda_conversion}
    \frac{f_{\nu}}{\rm [Jy]}=3.34\times 10^4 \left(\frac{\lambda}{\rm \AA}\right)^2 \frac{f_{\lambda}}{\rm [erg~cm^{-2}~s^{-1}~\AA^{-1}]},
\end{equation}
where $\lambda$ is the so-called pivot wavelength, that is, the measure of the effective wavelength of a filter. For ULTRASAT, we considered the average wavelength in its observational band ($\bar{\lambda}=260$~nm), while for the ZTF, we directly used values from the SVO Filter Profile Service\footnote{\url{http://svo2.cab.inta-csic.es/svo/theory/fps/}} \citep{svo1,svo2}. For both instruments, we obtain $f_{\lambda}^{\rm lim}\simeq 2\times 10^{-9}~\mathrm{erg~cm^{-3}~s^{-1}}$.

Figure~\ref{fig:evolution_time_comparison_bsg_rsg} shows the expected extinction-corrected $f_{\lambda}$ at different times after the SBO occurrence as a result of the outer stellar envelope cooling of RSG and BSG stars. The duration of the SBO can be roughly estimated as the progenitor light-crossing time  ($\Delta t_{\rm SBO} \simeq R_{*}/c$). For the set of fiducial parameters we adopted, $\Delta t_{\rm SBO} \sim 30(2)$ minutes for RGSs(BSGs). Here, a median redshift of $z=0.01$ (luminosity distance of about 40 Mpc) was set, in agreement with the redshift at which CCSNe are typically observed (e.g. \cite{median_z_ccsne}).
The figure also shows $f_{\lambda,\rm USAT}^{\rm lim}$, $f_{\lambda,\rm ZTF}^{\rm lim}$, and the instrument observation bands.
These results suggest that for representative RSGs and BSGs in an ambient medium with dust properties similar to those of our Galaxy, the UV emission from the SBO of either RSGs or BSGs would become visible to ULTRASAT approximately 5 minutes later. After at least 15(30) minutes, RSG(BSG) emission might also be detected by the ZTF. While the signal from RSGs remains visible in both optical and UV bands for around 6 months, the signal from BSGs is expected to last only $\sim$3 weeks($\sim$4 months), at which point it falls below the ULTRASAT(ZTF) detection threshold. Apparently, the much larger progenitor radius of RSG stars causes the UV and optical emission to remain widely visible for much longer as a result of the reduced cooling from expansion. 
Due to the strongly asymmetric $E_{B-V}$ distribution shown in Figure~\ref{fig:extinction}, 
the results obtained with the minimum and median extinction values are comparable. For the highest value of $E_{B-V}$ in the $2\sigma$ range, in contrast, (i) the emission from RSGs(BSGs) becomes visible $\sim$15(30) minutes after the SBO and (ii) BSGs remain visible to ULTRASAT for one week less.
Our results are consistent with the results from current instruments, which struggle to reveal CCSNe associated with BSG stars. In this regard, the role of ULTRASAT will be key for improving the capability of detecting this type of sources, at least for the closest ones. Unfortunately, as a side effect of the light absorption along the line of sight, the expected shape of the emission is more attenuated in correspondence of the ULTRASAT observational band because of the bump at 220~nm that is characteristic of the extinction curve \citep{cardelli}, as is well visible in the spectral dips in Figure~\ref{fig:evolution_time_comparison_bsg_rsg}.

We remark that we focussed on the UV and optical emission signal occurring on a timescale of days after the SBO. In principle, similar investigations can be conducted for the UV detectability of the SBO flash that forms when the shock reaches the edge of the star. This signal would be visible as a pronounced peak in X-ray/UV wavelengths right after the SBO occurrence on timescales ranging from seconds to a fraction of an hour, depending on the stellar envelope extension, as also pointed out beforehand.
However, as for ULTRASAT the available limiting magnitude only refers to sources with a colour temperature of $T_{\rm col}=20,000$~K, which does not correctly describe the early emission phases, we postpone the investigation of X-ray/UV flashes to a future study. For timescales shorter than one day, the colour temperature is related to the photospheric temperature through $T_{\rm col}\simeq 1.2 T_{\rm ph}$ \citep{rw11}, such that behind the shock wave, $T_{\rm col}$ might reach $\sim 10^5-10^6$ K.

\subsection{Computation of event rates}
\label{sec:nevents}
Following the arguments discussed in Section~\ref{sec:rate}, we calculated  the event rates per year $\dot{N}$ that is expected to be detected into the instrument FoV as a function of the time $t$ after the SBO occurrence, as
\begin{equation}
\label{eq:nevents}
\forall~ t \rightarrow \dot{N}=\int d\Omega \int_{0}^{z_{\mathrm{lim}}}\frac{dN(z)}{dz}dz=\int d\Omega \int_{0}^{z_{\mathrm{lim}}}\frac{R(z)}{1+z}\frac{dV(z)}{dz}dz,
\end{equation}
where $z_{\rm lim}$ is the maximum $z$ at which the model expected $f_{\lambda} > f_{\lambda}^{\rm lim}$ at given $t$. In the previous equation, $dV(z)/dz$ is the differential comoving volume, $R(z)$ is the comoving rate of sources, and $1+z$ comes from the cosmological time dilation of the observed rate. The comoving rate can be expressed as
\begin{equation}
    R(z)=R_0\frac{(1+z)^{2.7}}{1+\left ( \frac{1+z}{2.9} \right )^{5.6}},
\end{equation}
being $R_0=0.34 \times 10^{-4}$ Mpc$^{-3}$ yr$^{-1}$ the estimated local rate of H-rich SNe II without interaction with the CSM, computed as 60\% of the total rate of SN II \citep{li2011,lin_universe2023}. The redshift dependence follows the star formation rate from \cite{madau}.
We adopted the following assumptions: \\
(i) Because SNe II typically result from both RSG and BSG stars, while the ratio of BSGs and RSGs in our Galaxy is estimated to be $\rm BSG/RSG\simeq 3$ on average \citep{MW_metallicity}, we adopted as local rate for RSGs(BSGs) $R_0=1.1(2.5)\times 10^{-5}$ Mpc$^{-3}$ yr$^{-1}$. \\
(ii) In order to directly compare detector performances, we set the observation time of each instrument to one hour. In this time, the ZTF can scan 3750 deg$^2$ of the sky \citep{ztf}, while for ULTRASAT, we conceive an observational strategy of $4\times900~\rm s$, each covering the ULTRASAT $\rm FoV=204~deg^2$ \citep{ultrasat}. However, the source detection efficiency also depends on the instrument duty cycle. For ULTRASAT, this amounts to $100\%$, while for the ZTF, we considered just $25\%$ of the rate of detectable sources because the ZTF can only operate nightly and in good weather conditions \citep{ztf_observing_system}.
To characterise RSG and BSG populations, we assumed the same progenitor parameters as in Section~\ref{sec:emissions_sbo} with a median extinction equal to $E_{B-V}=0.04$ because this value characterises the SN II source population best. Equation~\eqref{eq:nevents} implicitly assumes that all that SNe II enter the calculation are characterised by the same $E_{B-V}$ ($z_{\lim}$ is calculated for a fixed $E_{B-V}$ value).

\begin{figure*}[t!]
\subfigure[\label{fig:zlim}]{\includegraphics[width=1.01\columnwidth]{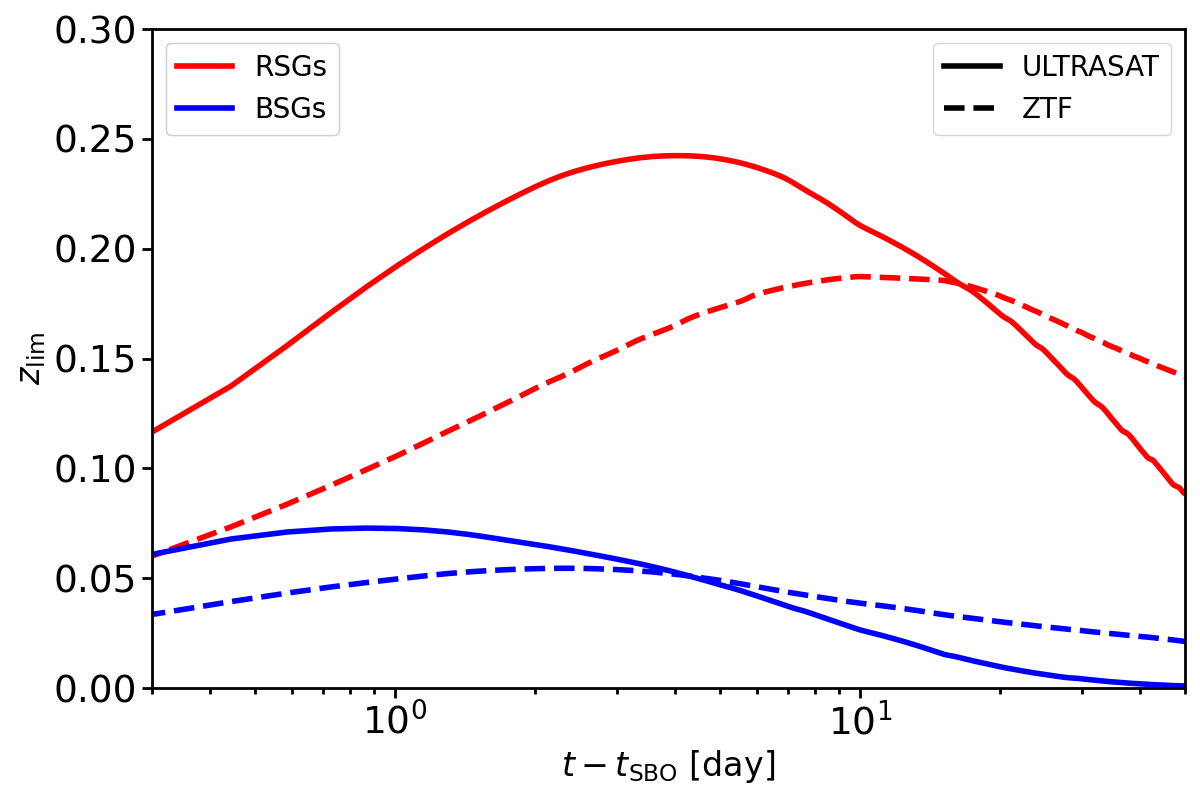}}
\subfigure[\label{fig:nevents}]{\includegraphics[width=1\columnwidth]{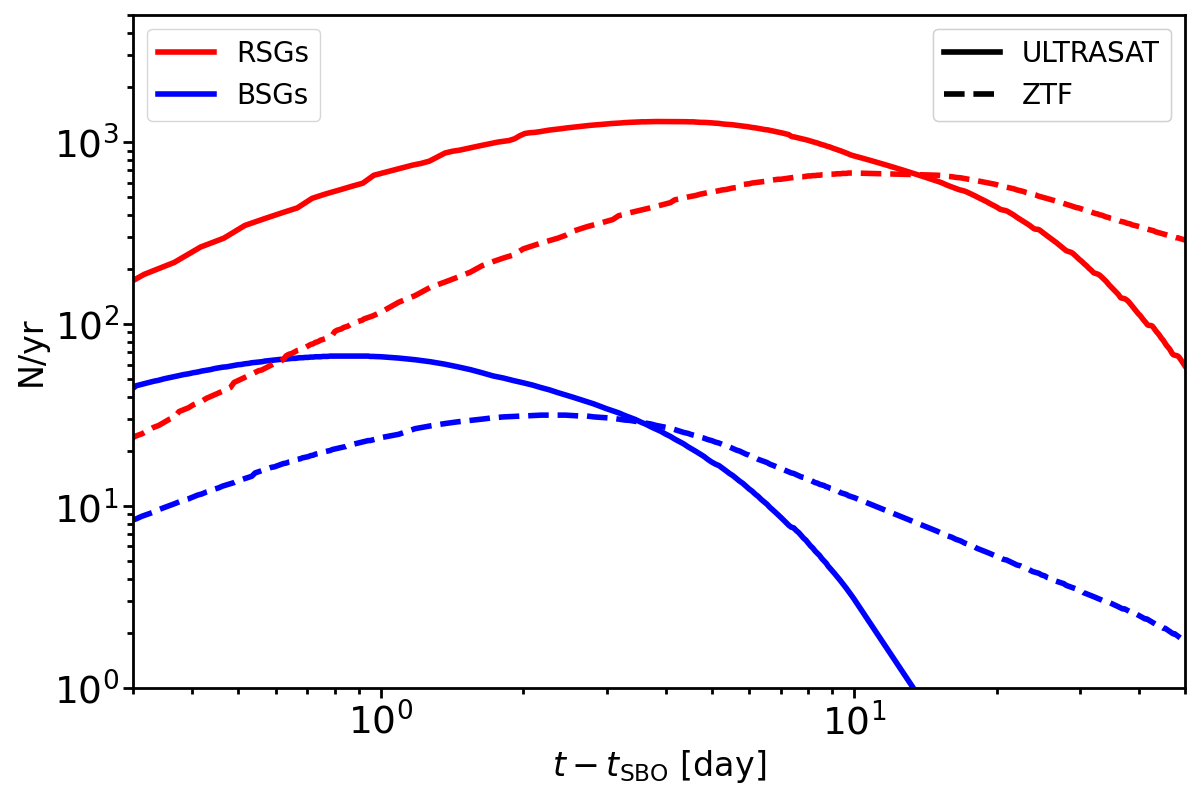}}
\caption{Capability of detecting UV and optical emission from RSGs and BSGs with ULTRASAT and ZTF. (a) Maximum redshift at which RSGs and BSGs would have UV and optical emissions detectable by ULTRASAT and ZTF as a function of the emission time, $t$. (b) Rate of SNe II per year from RSGs and BSGs detectable with ULTRASAT and ZTF as a function of the emission time, $t$. In both panels, the following fiducial parameters are adopted: for RSGs $R_{*}=722~R_{\odot}$, $E=10^{51}~\rm erg$, $M_{\rm ej}=2.8~M_{\odot}$, and $f_{\rho}=1.455$; for BSGs $R_{*}=50~R_{\odot}$, $M_{\rm ej}=10~M_{\odot}$, $f_{\rho}=0.0465$, and $E=10^{51}~\rm erg$. The emission time corresponds to the time elapsed since the SBO. Results for RSGs and BSGs are shown in red and blue, respectively. The solid and dashed lines show ULTRASAT and ZTF results, respectively.}
\label{fig:zlim_nevents}
\end{figure*}

As a result, Figure~\ref{fig:zlim} shows the maximum redshift as a function of the elapsed time since the SBO from which the signal can be revealed by ULTRASAT and ZTF. The trend shows a growing horizon for a few days after the SBO (depending on the observing instrument), which later decreases and which reflects a visibility peak in the emission. At later times, only closer sources can be still accessible. We also note that ULTRASAT will exploit larger volumes of the Universe for these explosions and will be able to catch observations from farther redshifts.
Figure~\ref{fig:nevents} shows the corresponding number of SNe II per year detectable by ULTRASAT and the ZTF (following Equation~\eqref{eq:nevents}) under these assumptions.

\section{Results and implications for multi-messenger observations}
\label{sec:mm_followups}
In this section, we discuss our results within the context of multi-messenger observations and focus at first on UV and optical electromagnetic signals visible by ULTRASAT and ZTF in Section~\ref{sec:em_signals} and later on neutrino associations in Section~\ref{sec:neutrinos_implications}.

\subsection{Ultraviolet and optical follow-ups}
\label{sec:em_signals}
Figure~\ref{fig:zlim} enables us to derive the following conclusions:

(i) For RSG progenitors, a UV (optical) signal from SBOs by ULTRASAT (ZTF) can be detected out to $z\sim0.24 (0.19)$, corresponding to luminosity distances of $\sim 1.2~\rm Gpc~(920~\rm Mpc)$ when the SBO occurred $\sim4(10)$ days before detections. Later emissions, as well as signals from times closer to the SBO, can still exceed the sensitivity of the detectors only for closer SNe II. 

ii) For BSG progenitors, the probability of detecting an analogous signal is lower than RSGs because of the different stellar envelopes, which cause a weaker emission, as discussed in Section~\ref{sec:rate}. In particular, ULTRASAT (ZTF) can detect SBO signals from BSGs located up to $z\sim0.07(0.05)$, that is, from luminosity distances of $\sim 310(220)~\rm Mpc$. 
About three weeks after the SBO, the signal from BSGs is no longer detectable by both detectors, whereas the signal from RSGs remains detectable for six months.
Furthermore, we highlight that our predictions about the maximum redshift reachable by the ZTF are consistent with published results about the ZTF SN sample, whose public survey probes SNe II out to approximately $z<0.05$ \citep{ztf_ccsn}. 

These results are expected to help the scientific community in constraining the progenitor of SNe II observed in both UV and optical wavelengths via ULTRASAT and ZTF-like instruments, as well as in identifying possible choked jets taking advantage of a multi-wavelength strategy. When the UV and optical emission lasts more than ten days, this may indicate that 
the progenitor was a RSG star. Moreover, the lack of a $\gamma$-ray counterpart (i.e. the associated GRB) indicates that the observed emission stemmed from a choked jet because successful GRBs exploded at the distances discussed here are typically detected by current instruments.

To firmly confirm the hypothesis of hidden jets harboured in stellar envelopes, additional follow-ups are needed. A few observations might serve as evidence, such as (i) very broad absorption features in SN early spectra because of the fast cocoon material that engulfs the star when the hot cocoon material breaks out and spreads, and (ii) a high-velocity component in the energy-velocity profile of SNe \citep{piran2019,Nakar2015,pais2023}. Different spectra taken at different times should allow the SN velocity profile to be compared to the regular spherical explosion case. During the first few days, the cocoon material that has spilled around the star produces very broad absorption features because of its high-velocity expansion in an optically thick environment. After a few days, when this becomes transparent, the very broad lines of the cocoon disappear. At this stage, the SN ejecta dominates the absorption. Many stripped-envelope SNe have shown these features, which supports the hypothesis that hidden jets are connected to CCSNe. This was also the procedure adopted to probe the possible presence of a relativistic choked jet in GRB 060218/\textit{SN 2006aj} \citep{Nakar2015}. However, these types of investigations may not be straightforward for massive H envelopes such as those we considered. These envelopes might choke not only the jet, but partially the cocoon as well, thereby making the observations of these signatures from the early spectra demanding.

Missing $\gamma$ rays can also result from off-axis GRBs that despite the presence of a successful jet do not point directly towards Earth. Their beamed emission would therefore not be detected. However, their SBO emission is expected to last much shorter than what we considered here. For large viewing angle models, the SBO luminosity suddenly reaches a peak value and declines. In other words, if UV and optical emissions are detected over several days, it is reasonable to think that such signals are not related to off-axis GRBs \citep{suzuki2010}.

Figure~\ref{fig:nevents} shows the expected rate of detectable SNe II as a function of time after the SBO event. After ULTRASAT is operational, the multi-messenger and multi-wavelength community can clearly profit from at least as many observations as in the optical field, or even more. Under the present assumptions, ULTRASAT will be able to reveal up to $\sim$1300(65) SNe II per year from RSGs(BSGs) if the detection is performed within $\sim$4(1) day(s) after the SBO. About $50\%$ of SNe II from RSGs can also be accompanied by an optical detection by the ZTF if it catches optical emission within $\sim$10 days after the SBO (i.e. around one week after the UV emission). As already pointed out in Figure~\ref{fig:zlim}, the ZTF cannot perform sky observations as deep as ULTRASAT.
As regards to the signals produced by BSG progenitors, one source out of the three that will be detectable by ULTRASAT per year might be associated with optical measurements by ZTF. 

Note that the ULTRASAT $5\sigma$ mean limiting magnitude, $m_{\mathrm{AB}}^{\rm lim,USAT} = 22.5$, corresponds to a 900 s integration time. However, the nominal exposure time for individual observations is 300 s \citep{ultrasat}. The decision on whether transient alerts will be issued based on individual 300 s exposures or coadded stacks (i.e., $3 \times 300$ s) has not yet been finalised. We estimate the limiting magnitude for a 300 s exposure as $m_{\rm AB}(300~\mathrm{s})=m_{\rm AB}(900~\mathrm{s})-1.25\log_{10}(900/300)=21.8$, valid under the assumptions of background-limited observations and Gaussian noise. This corresponds to a $\sim 0.7$ magnitude reduction in sensitivity compared to stacked 900 s exposures, leading to an increase in the limiting flux by a factor of $\sim 1.7$. Consequently, the maximum redshift for different progenitor types decreases; for example, for RSGs (BSGs), the maximum redshift drops from 0.25 (0.07) to 0.2 (0.06). On the other hand, 300 s exposures enable ULTRASAT to cover three times the sky area per unit time. Qualitatively, using the same calculations as for 900 s exposures and assuming the same FoV of $204$ deg$^{2}$, this implies an increase of up to approximately 35\% in the number of observed sources. However, this gain comes at the cost of reduced sensitivity, resulting in fewer detections of faint and distant sources.

The results presented here can be influenced by specifics observational survey strategies implemented by the two instruments. Any real survey does not cover the whole sky, operates over a finite time window with a complex cadence structure, must contend with different values of Galactic extinction, and does not recover all the transients it is able to detect in principle because of experimental selection criteria. The limiting magnitudes that can be reached by detectors change with the exposure time (the higher the latter, the lower the minimum flux detectable by each instrument). Additionally, while ULTRASAT will observe a large patch of the sky in the southern and northern hemispheres, the ZTF scans declination values above $-30^{\circ}$ ($\sim3\pi$ of the sky). In the light of this, we encourage the optimisation of combined observations from different collaborations  for a well-defined follow-up strategy tuned upon the results presented here. 

Within the multi-messenger context, neutrinos can also play a crucial role. The interaction of accelerated protons and thermal photons and/or hadronic collisions in choked jets can lead to the production of $\nu$s that are able to escape from the thick stellar envelope of the system. This scenario recently attracted much attention (e.g. \citealt{MuraseIoka2013,Senno2016,choked1,choked3,fasano}) because it might explain the astrophysical diffuse flux (TeV-PeV neutrino energies), that was detected for the first time by the IceCube Neutrino Observatory (IceCube), without incurring inconsistencies with the isotropic diffuse $\gamma$-ray background observed by Fermi (e.g. \citealt{fermi_inconsistence}).
In the next section~\ref{sec:neutrinos_implications}, we discuss how neutrino observations can be combined with electromagnetic signals from SBO events.

\subsection{Neutrino follow-ups}
\label{sec:neutrinos_implications}
During stellar collapse, neutrinos can escape the thickest envelopes, such that their detection would constitute an early warning for the multi-messenger astronomical community, advertising that the light from the explosion is coming. This would trigger the search for the SBO electromagnetic signals following core collapse. Therefore, the combination of multi-messenger signals from UV, optical, and neutrino emissions provides an unprecedented opportunity to probe the existence of choked jets, and shed light on their progenitors.

High-energy neutrino telescopes, such as IceCube \citep{icecube} and the Cubic Kilometre Neutrino Telescope (KM3NeT) \citep{loi_km3net}, the latter of which is under construction at the bottom of the Mediterranean Sea and taking data in a partial detector configuration, may be able to reveal a flux of neutrinos from SNe II with choked jets, depending on their distance and energetics. Real-time analysis systems in these instruments allow for a prompt reaction, resulting in a distribution to the multi-messenger community  within a few seconds. The high duty cycles of these instruments, combined with the all-sky field of view, make them ideal partners in multi-messenger strategies.
IceCube has been sending neutrino alerts to external communities for triggering subsequent follow-ups since 2016 \citep{icecube_online}, while the plan for KM3NeT is to automatically start sending alerts within the end of this year (see \citealt{Celli:2023tG} for the description about the current status of the KM3NeT online system). For the reception of external triggers, KM3NeT has been reacting to any external multi-messenger alert since November 2022 via pipelines that automatically run analyses within time windows optimised to several classes of sources \citep{PalaciosGonzalez:202306}. Instead, IceCube uses a fast-response analysis (FRA) pipeline \citep{fra} that is manually activated to only follow up selected interesting astrophysical transients.

The idea of performing optical follow-up of single high-energy neutrinos to find SNe dates back several years ago \citep{neutrinos_optical_followup1}. We provide further indications for defining a proper strategy, focused on Type II CCSNe. We encourage UV, optical, and neutrino telescopes to optimise both their alert sending and external follow-up programmes based on the results presented in this manuscript. When the observational strategy that we propose is combined with photometric and spectroscopic studies, it would be crucial to unveil choked jets. By considering the maximum probability to detect SNe II (see Section~\ref{sec:mm_followups}), we can argue that when an interesting neutrino alert is released, ULTRASAT (ZTF) could point to the suggested direction of the sky within around 4 (10) days to search for possible electromagnetic counterparts, thus maximising the reachable sky volume and hence the number of detectable sources. However, this time window leads to the time of the SBO occurrence. To also consider the production of neutrinos during the shock-propagation time inside the stellar envelope, we need to enlarge this time window up to about one day.
In particular, we expect the shock wave to take about less than one day to propagate in the radiative envelopes of BSGs and about one day in the convective envelopes of RSGs for values of the explosion energy considered in this work \citep{kistler}.

It is worth stressing that neutrino and electromagnetic follow-ups are extremely important today. Although over the past years, the IceCube Collaboration has reported some observational indications of sources contributing to the diffuse neutrino flux, the majority of the diffuse flux is still unexplained. The blazar TXS 0506+056  and the nearby active galactic galaxy NGC 1068, which were identified as possible sources of high-energy neutrinos with a $3\sigma$ \citep{txs} and $4.2\sigma$ significance \citep{ngc1068}, respectively, are expected to contribute no more than $\sim1\%$ to the diffuse flux. In addition, strong evidence ($\sim4.5\sigma$) for neutrino emission from the Galactic plane was recently reported \citep{icecube_galacticplane}, with an estimated contribution to the diffuse flux at 30 TeV of $\sim 10\%$ \citep{icecube_galacticplane}. Hence, most of the observed emission remains unexplained, and many other astrophysical sources are needed to account for the remaining part.

Recently, CCSNe have been tested as potential sources of the diffuse neutrino flux in the energy range of about $10^3-10^5$ GeV, assuming that the neutrino energy spectrum follows a power law with an index of -2.5 \citep{icecube_ccsne}. Correlations between seven years of IceCube neutrino data and a catalogue containing more than 1000 CCSNe of types IIn and IIP and a sample of stripped-envelope SNe have been searched for via either individual source studies and stacking analysis for combined emission from the whole sample. Even though no significant spatial and temporal correlation between CCSNe and neutrinos was found, this study ruled out CCSNe of type IIn and stripped-envelope SNe as the dominant source of the diffuse neutrino flux. Current limits indicate that type IIP SNe (associated with RSGs, as discussed in Section~\ref{sec:introduction}) might at most contribute $\sim 60\%$ of the diffuse flux (under the aforementioned assumptions) at most to the production of high-energy neutrinos.
This strongly motivates the follow-ups that we propose here because we focus on the search for high-energy neutrinos from H-rich type II CCSNe, to which the RSG class belongs.

With regard to the number of combined detections that are expected to include neutrinos, we expect the rate of these sources to be at the percent level of the RSG local rate when we limit the high-energy neutrino production to RSGs hosting choked jets. This value results from considering the number of choked jets required to explain current IceCube constraints on CCSNe as follows. \citealt{fasano} computed the diffuse neutrino flux from a population of choked jets from RSGs with stellar progenitor radius of $\sim 500R_{\odot}$ through detailed calculations of $p \gamma$ interactions, and compared it to the entirety of the detected diffuse astrophysical neutrino flux, as at the time the contributions from the Galactic plane and active galactic nuclei did not yet emerge. The comparison allowed us to provide constraints on the local rate of this source population to reproduce observations: For GRB energies channelled into protons between $10^{51}$ and $10^{53}$ erg, the local rate of choked jets required to reproduce the observed astrophysical neutrino flux was between 80 and 1 Gpc$^{-3}$ yr$^{-1}$, respectively. 
Given the CCSN limit set by IceCube, we should restrict this rate to 60\% for a more realistic estimate of choked jets, corresponding to a fraction of $\sim$0.5\% of the estimated local rate of RSGs adopted here (i.e., $1\times 10^{4}$Gpc$^{-3}$ yr$^{-1}$; see Section\ref{sec:rate}).
Hence, of the $\sim$1300($\sim$650) RSGs per year detectable in the UV(UV+optical) band by ULTRASAT(ULTRASAT+ZTF) within 1.2 Gpc, approximately 20\% could be sources hosting choked jets that produce high-energy neutrino emissions. These emissions could contribute to explaining the origin of the cosmic diffuse neutrino flux. The local rate of choked jets required to match the observed astrophysical neutrino flux from \cite{fasano}, rescaled by the CCSN limit set by IceCube, could be achieved after approximately four years. This evaluation shows the essential necessity to run multi-messenger analyses like this for several years. 

The ZTF is already performing a systematic neutrino follow-up program to search for optical counterparts of high-energy neutrinos using different search-time windows up to 14 days after each IceCube neutrino alert. \citealt{stein_ztf} presented results from the first 24 campaigns. Since March 2018, the TF has taken prompt observations for 24 high-quality IceCube neutrino alerts, with a median latency of 12.2 hours from initial neutrino detection. Optical signatures from Type IIn SNe, SNe with relativistic jets, GRB afterglows, and active galactic nuclei flares have been considered, without identifying additional candidate neutrino sources. As stressed by the authors themselves, the strategy presented in \citealt{stein_ztf} can serve as a pathfinder for future triggered neutrino follow-up programs with wide-field instruments.
Furthermore, other neutrino follow-up programs in the optical wavelengths are also in place thanks to the emergence of different surveys over the past years, such as the Panoramic Survey Telescope and Rapid Response System (Pan-STARRS) \citep{panstarrs,panstarrs_sn}, the All-Sky Automated Survey for Supernovae (ASAS-SN) \citep{asassn,asassn_sn}, and the Dark Energy Camera (DECam) \citep{decam,decam_sn}. Future UV detections with ULTRASAT will complete the current picture, enabling the multi-messenger community to study the dynamic neutrino sky across the entire electromagnetic spectrum.

It is also worth pointing out that the approach presented in this paper (described in Section~\ref{sec:detection_sbo}) can be extended to combined searches between any existent electromagnetic facility and neutrino telescopes. When a SN is detected somewhere in the Universe in a UV and/or optical wavelength band, it is possible to define a reasonable time window when looking for neutrinos coincident with different emission phases of that SN by knowing its host galaxy (hence, its distance and the characteristic extinction) and by combining the information about the expected flux by the model and the limit flux detectable by the electromagnetic instrument reporting the detection. For instance, the IceCube Collaboration has recently adopted this procedure in a FRA run with the aim of performing a search for track-like muon neutrino events arriving from the direction of the nearby Type IIn SN 2024bch \citep{sn2024bch_1,sn2024bch_2} over two different time windows in order to detect emission from either the shock-propagation wave inside the stellar progenitor up to the shock breakout or the interaction of the ejecta with the CSM \citep{zegarelli_atel}. Because of no detections, upper limits on neutrino production from SN 2024bch were derived, under the simple assumption that a neutrino energy spectrum follows an $E^{-2}$ power law, as traditionally predicted by the Fermi shock acceleration \citep{fermi_acceleration}.

Finally, our results can also be used to more accurately tune the neutrino search-time window of offline analyses performed in correlation with observed SNe \citep{senno2018,stacking_choked}. Offline comprehensive studies might even provide higher significance than real-time analyses via stacking sources from catalogues of optical and UV emitters.

\section{Summary and conclusions}
\label{sec:conclusions}
We investigated the radiative emission from core-collapse SNe with extended H envelopes (SNe II), specifically, RSGs and BSGs. These sources are particularly interesting because their lack of GRB counterparts indicates that these SN types can potentially harbour jets that are choked within the stellar envelopes. In the past decade, a few indications of jets in SNe have been reported. From the point of view of neutrinos, choked jets are fascinating because they might contribute to the astrophysical diffuse flux detected by IceCube. 

Following the SBO occurrence, the stellar envelope expands and cools nearly adiabatically. As the photosphere penetrates the outer shells of the envelope, radiation is produced in the UV and optical band. We focussed on this radiative signal and investigated future multi-messenger prospects for combined UV, optical and neutrino observations between the future UV satellite ULTRASAT, the currently operating optical telescope ZTF, and high-energy neutrino telescopes such as IceCube and KM3NeT.
We followed the model by \cite{rw11} to estimate the expected photon flux by considering fiducial values for stellar progenitor parameters: $R_{*}=722~R_{\odot}$, $E=10^{51}~\rm erg$, $M_{\rm ej}=2.8~M_{\odot}$, and $f_{\rho}=1.455$ for RSGs, and $R_{*}=50~R_{\odot}$, $E=10^{51}~\rm erg$, $M_{\rm ej}=10~M_{\odot}$, and $f_{\rho}=0.0465$ for BSGs. 
We then evaluated the possibility of detecting the extinction-corrected signals, taking advantage of ULTRASAT for the UV band at (230-290)~nm and the optical instrument ZTF. In this way, we characterised the future detection prospects of these sources, which we summarise as follows:
\begin{itemize}
    \item The farthest distance out to which a UV (optical) signal produced by the cooling emission after SBO in RSGs can be detected by ULTRASAT (ZTF) is $z\sim0.24$$(0.19)$, or $\sim 1.2~\rm Gpc$$(920~\rm Mpc)$, if the SBO occurred $\sim4$$(10)$ days before detections. Lower-redshift SNe can still produce signals exceeding the limiting magnitude flux of the detectors, even from later emission times or at times closer to the SBO. 
    \item The probability of detecting a similar signal from BSGs is lower than in the case of RSGs because of the different characteristics among the two progenitors and subsequent less pronounced emission resulting from SBO in BSGs, as discussed in Section~\ref{sec:rate}. In particular, ULTRASAT (ZTF) can detect signals from an SBO that occurred in BSGs located up to $z\sim0.07$$(0.05)$, or $\sim 310$$(220)~\rm Mpc$.
    \item ULTRASAT will be able to reveal up to 1300(65) SNe II per year from RSGs(BSGs) if the detection is performed within $\sim4$(1) days after the SBO.
    \item Around $50\%$ of SNe II from RSGs can also be accompanied by an optical detection by the ZTF, if it catches optical emission within $\sim$10 days after the SBO (i.e. about one week after the UV detection).
    \item One source out of the three detectable by ULTRASAT in one year in the case of SNe II from BSGs may be associated with optical measurement by the ZTF.
\end{itemize}

Although our findings suggest that ZTF’s maximum detectable CCSN rate exceeds its actual detection capacity, this discrepancy arises because our prediction assumes that all SNe are observed about 10 days after shock breakout, when their emission peaks according to our model. In reality, this idealized scenario does not fully reflect observational conditions, where SNe are typically detected within a few days.
More precisely, ZTF detects approximately 300 CCSNe per year, with ~180 being Type II-P and II-L, the focus of this work. This sample primarily includes the most luminous events, while $\sim$200 additional SNe remain “unclassified” due to their low luminosity, preventing accurate classification. According to \cite{perley}, Type II SNe account for $\sim$20\% of this sample (see Figure 6). Even considering this, ZTF is estimated to detect no more than $\sim$200 CCSNe of interest annually. As shown in Figure 3(b), this aligns with current surveys, where ZTF detects up to $\sim$200 CCSNe per year when observations occur within a few days of the event.
Furthermore, we adopt the $5\sigma$ median sensitivity across all filters ($g$, $r$, and $i$ bands) and lunar phases for a 30-second exposure, yielding $m_{\rm AB}^{\rm lim,ZTF}=20.4$ \citep{ztf} (see Section~\ref{sec:detection_sbo}). However, publicly available TNS classifications (\url{https://sites.astro.caltech.edu/ztf/bts}), used to derive the above estimates, indicate that detected SN events often have magnitudes below 19. This is likely due to selection cuts or data limitations that exclude sources based on criteria such as luminosity or signal-to-noise ratio.

As these sources can also produce neutrinos via interactions between protons and thermal photons in the choked jets, neutrino observations by existing Cherenkov high-energy neutrino telescopes (e.g. IceCube and KM3NeT) can be used in association with electromagnetic signals from SBO events. Both IceCube and KM3NeT have leading roles in the multi-messenger community, working in synergy with several partners, in both offline and online analyses. In particular, much effort has been devoted to reconstructing and classifying their own data in real time to alert external communities to interesting neutrino events and to follow-up interesting astrophysical transients revealed by other facilities. We find that
\begin{itemize}
    \item by considering the maximum probability of detecting SNe II, ULTRASAT (ZTF) could point to the provided direction of the sky within $\sim4$ (10) days to search for possible electromagnetic counterparts when an interesting neutrino alert is released, under the hypothesis that it comes from the SBO emission phase. In this way, the sky volume that can be reached and hence the number of detectable sources is maximised.
    \item to consider the possibility that neutrinos are produced during the shock propagation time inside the stellar envelope, electromagnetic observations can wait up to one day longer with respect to the times indicated above.
\end{itemize}

We find that UV, optical, and neutrino follow-ups adopting the strategy that we present here, combined with photometric and spectroscopic studies, would be crucial to unveil choked jets. By considering the recent constraints on the local rate of choked jets from RSGs computed by \citet{fasano} to reproduce the IceCube diffuse flux, and restricting their contribution to 60\% of the rate derived there at most, as indicated by the upper limit set by IceCube Collaboration in CCSN searches \citep{icecube_ccsne} as compared to the estimated local rate of RSGs, we find that up to approximately 20\% of detectable CCSNe from RSGs can host a choked jet that produces high-energy (>TeV) neutrinos. Since ULTRASAT is expected to detect $\sim1300$ RSGs per year in the UV band (and ULTRASAT+ZTF will detect $\sim650$ in UV+optical) within 1.2 Gpc, conducting UV, optical, and neutrino follow-ups over several years will be essential to definitively constrain the contribution of RSGs to the cosmic diffuse neutrino flux. This would enable us to determine whether choked jets are a dominant source of multi-TeV neutrinos and thereby explain the origin of the high-energy diffuse neutrino flux.

In addition to fast follow-ups on single serendipitous sources (discussed in Section~\ref{sec:neutrinos_implications}), the considerations presented here can also be used for offline analyses based on the comparison between SN catalogues and neutrino data. In these studies, the proper definition of the time window in which searching for neutrinos is fundamental. The results discussed here allow a tighter constraint of the connection between a choked jet from SNe and high-energy cosmic neutrinos in the near future compared to studies in the literature (e.g. \citealt{senno2018,stacking_choked}).

As already pointed out throughout the paper, at the moment of writing, the end of the current ZTF program is uncertain, and it is still unknown whether other runs in the future  might overlap the ULTRASAT activity (whose launch is expected in 2026). Hence, all the results we discussed here are to be interpreted in terms of ZTF-like instruments. Interestingly, the Large Array Survey Telescope (LAST), a multipurpose telescope array in construction in the Israeli Negev Desert, will provide great support to ULTRASAT by monitoring the northern sky and providing simultaneous data in the visible wave band \citep{last1,last2}. LAST currently includes 32 out of 48 telescopes, and the deployment of the remaining telescopes is planned for the near future. 
In the future, optical surveys will also be complemented with measurements by the \textit{Vera C. Rubin} Observatory in Chile, which will observe the southern sky \citep{lsst}. Preliminary estimates of the expected number of \textit{Rubin} detections for different SN progenitor types have recently been published in \cite{lsst_estimations}.
Even though \textit{Rubin} will be more sensitive than LAST (see Figure 1 in \citealt{last2}, where the relative volume per unit time visible by several sky surveys is shown), LAST can provide follow-up observations for alerts that are not immediately observable by the ZTF or \textit{Rubin} due to its geographic locations.
These future multi-messenger campaigns can benefit from the results presented here to define their observational strategies.

\begin{acknowledgements}
We first would like to thank the anonymous referee who provided useful and detailed comments on the first version of the manuscript. We acknowledge the support of the Sapienza School for Advanced Studies (SSAS) and the support of the Sapienza Grants No. RM120172AEF49A82, RG12117A87956C66 and RM1221816813FFA3. We gratefully acknowledge Antonio Capone for the useful discussions and comments to our work. We also thank Brad Cenko and Anna Franckowiak for the fruitful conversations and useful insights on the topics treated in the present paper. Finally, we would like to thank also Robert Stein for his useful comments and suggestions. This work made use of Astropy:\footnote{http://www.astropy.org} a community-developed core Python package and an ecosystem of tools and resources for astronomy \citep{astropy:2013, astropy:2018, astropy:2022}.
\end{acknowledgements}

%%%%%%%%%%%%%%%%%%%%%%%%%%%%%%%%%%%%%%%%%%%%%%%%%%%%
\bibliographystyle{aa} % style aa.bst
\bibliography{aanda.bib} % your references Yourfile.bib

\begin{appendix}
\section{Impact of the parameter space of the progenitors on the expected signals}
\label{appendix:uncertainties}

\begin{figure}
\centering
\subfigure[\label{fig:radius_uncertainty_RSG}]{\includegraphics[width=1\columnwidth]{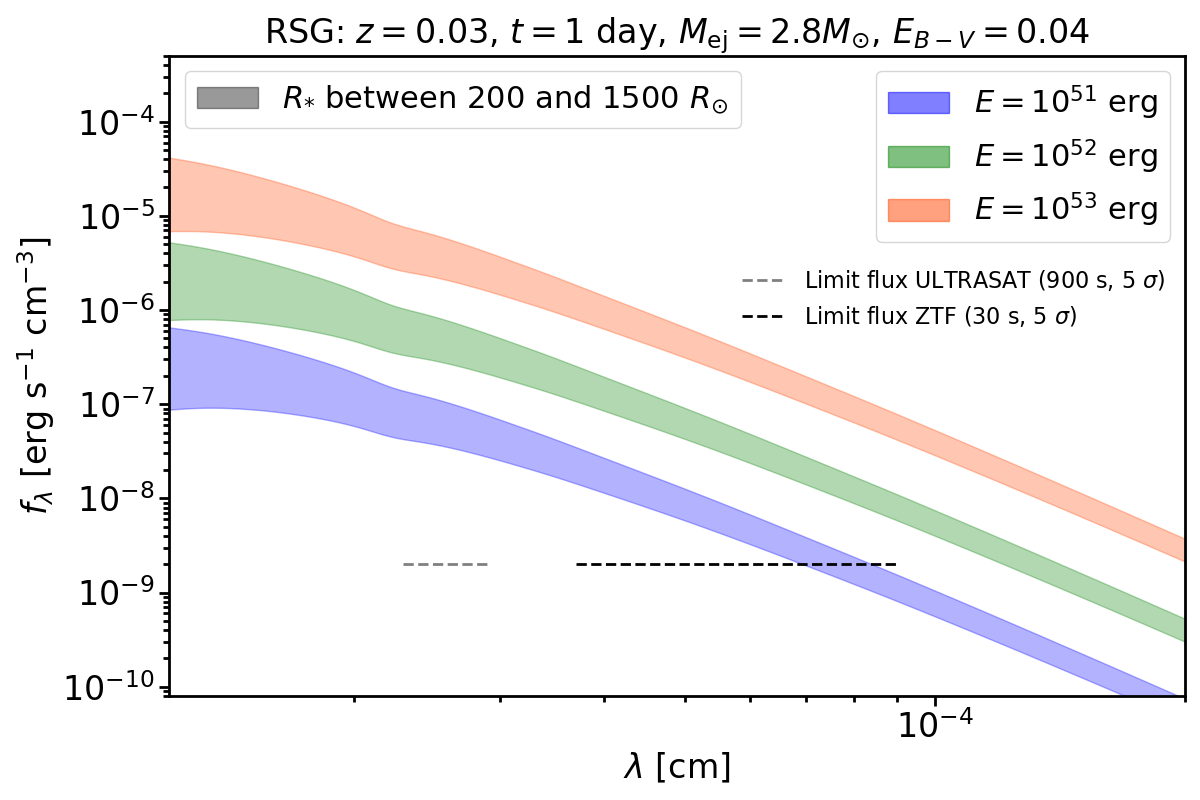}}
\subfigure[\label{fig:radius_uncertainty_BSG}]{\includegraphics[width=1\columnwidth]{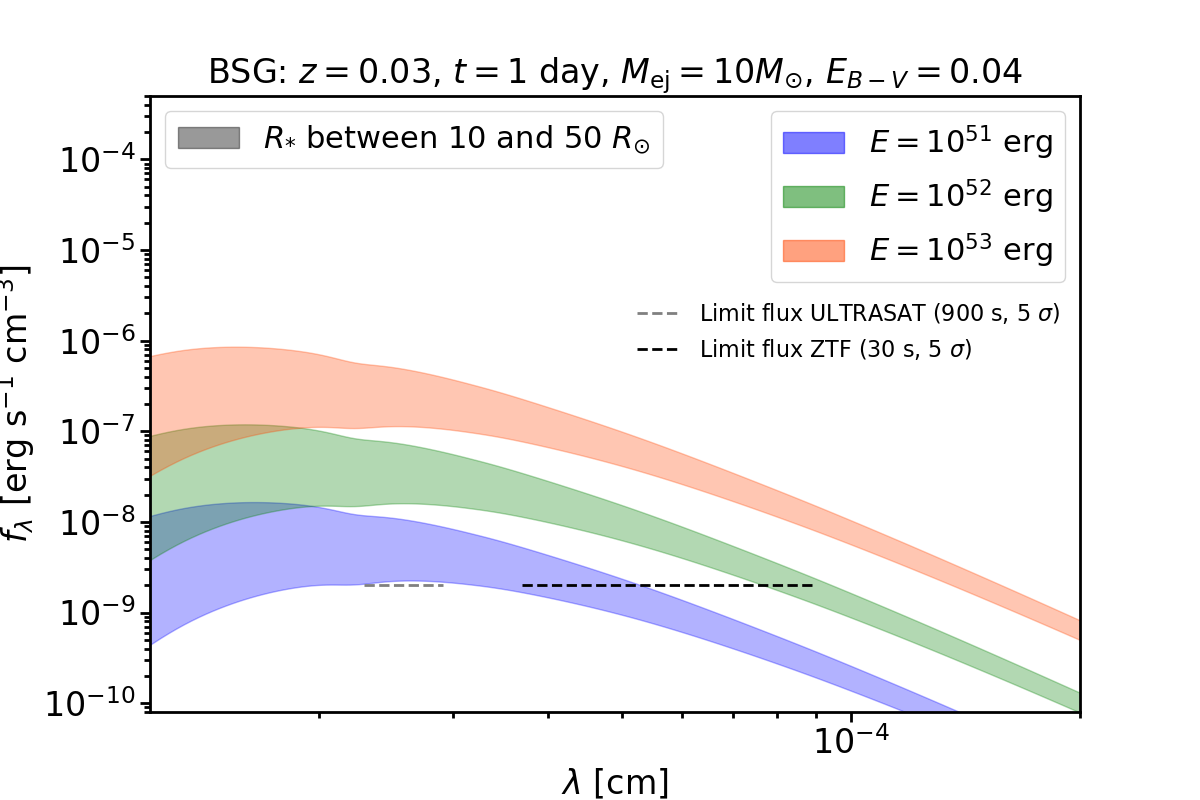}}
\caption{Model specific intensities $f_{\lambda}$ (in Equation~\eqref{eq:flambda}) observed at Earth from RSGs (a) and BSGs (b) located at $z=0.03$ hosted in a galaxy with extinction value $E_{B-V}$ (median value from Figure~\ref{fig:extinction}). 
Blue, green, and orange shaded regions show results for explosion energy $E=10^{51}$~erg, $E=10^{52}$~erg, and $E=10^{53}$~erg, respectively. (a) RSG progenitor: by fixing the ejected mass at $M_{\rm ej}=2.8~M_{\odot}$ (fiducial parameter), the shaded regions reflect the variation of the progenitor radius between 200 and 1500 $R_{\odot}$. (b) BSG progenitor: by fixing the ejected mass at $M_{\rm ej}=10~M_{\odot}$ (fiducial parameter), the shaded regions reflect the variation of the progenitor radius between 10 and 50 $R_{\odot}$. $f_{\lambda}^{\rm lim}$ of ULTRASAT and ZTF in their observational wavelength bands, obtained following Equation~\eqref{eq:fnu_flambda_conversion}, are indicated with the grey and black dashed horizontal lines, respectively.}
\label{fig:radius_uncertainty}
\end{figure}
In the present Section, we investigate how the radiative signal fluxes considered in this work are affected by the variation of parameters that characterises the stellar progenitor. To this aim, we fix the value of Galactic extinction to $E_{B-V}=0.04$, corresponding to the median extinction value for galaxies hosting SN Type II, as shown in Figure~\ref{fig:extinction}, the time of detection to 1 day after the SBO, and the redshift to $z=0.03$.
As discussed in Section~\ref{sec:emissions_sbo}, the radius and the effective temperature of the photosphere penetrating into the outer shells of the envelope in choked-jet systems directly influence the emissivity $f_{\lambda}$ arising after the SBO (see Equation~\eqref{eq:flambda}), and these in turn are determined by stellar progenitor characteristics, namely by the stellar radius $R_{*}$, the explosion energy $E$, the ejected mass $M_{\mathrm{ej}}$. 

\begin{figure}
    \centering\includegraphics[width=1\columnwidth]{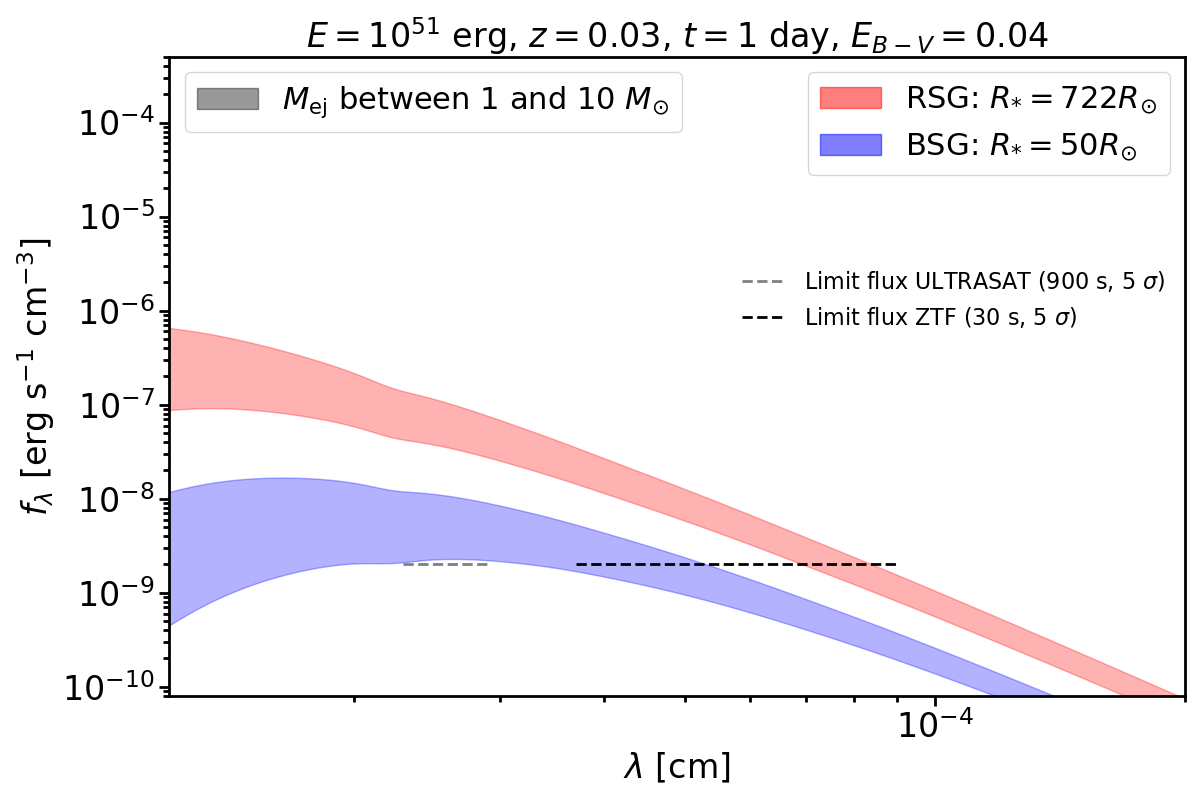}
    \caption{Model specific intensities $f_{\lambda}$ (in Equation~\eqref{eq:flambda}) observed at Earth from RSGs (in red) and BSGs (in blue) located at $z=0.03$ hosted in a galaxy with extinction value $E_{B-V}$ (median value from Figure~\ref{fig:extinction}).
    For each progenitor type, fiducial parameters are used (for RSGs $R_{*}=722~R_{\odot}$, $E=10^{51}~\rm erg$ and $f_{\rho}=1.455$; for BSGs $R_{*}=50~R_{\odot}$, $f_{\rho}=0.0465$, and $E=10^{51}~\rm erg$), apart from the ejected mass $M_{\rm ej} $that is let to vary between 1 and 10 $M_{\odot}$. $f_{\lambda}^{\rm lim}$ of ULTRASAT and ZTF in their observational wavelength bands, obtained following Equation~\eqref{eq:fnu_flambda_conversion}, are indicated with the grey and black dashed horizontal lines, respectively.}
    \label{fig:mass_uncertainty}
\end{figure}

Throughout the work, we used fiducial parameters for both RSG and BSG progenitors. Here, we vary the stellar radii over ranges of possible parameters, namely [200-1500]~$R_{\odot}$ and [10-50]~$R_{\odot}$ for RSGs and BSGs, respectively, the ejecta mass values between 1 and 10 $M_{\odot}$, and we also consider energies between $10^{51}$ and $10^{53}$ erg. Figure~\ref{fig:radius_uncertainty} reports $f_{\lambda}$ for different $E$ and $R_{*}$ within the aforementioned range of values. In particular, the left and right panel refer to RSG and BSG progenitors, respectively.
In such a case, we adopt the fiducial $M_{\mathrm{ej}}$ values used previously in the manuscript. Clearly, (i) the higher the explosion and the larger the emission arising, and (ii) the more extended the stellar envelope and the higher the level of $f_{\lambda}$, because of the increasing amount of shocked material impacting the outer shell of the progenitor. This is more evident for RSG progenitors where $r_{\mathrm{ph}}\propto E_{51}^{0.41}$ and $T_{\mathrm{ph}}\propto E_{51}^{0.027}$, while milder dependencies a $r_{\mathrm{ph}}\propto E_{51}^{0.39}$ and $T_{\mathrm{ph}}\propto E_{51}^{0.016}$ characterise BSGs (see Equation~\eqref{eq:phradius} and Equation~\eqref{eq:tph}).
In Figure~\ref{fig:mass_uncertainty}, we fix the fiducial parameters for RSGs and BSGs, allowing the ejecta mass to vary between 1 and 10 solar masses: in such a case, $f_{\lambda}$ is affected by less than a factor 10.

\end{appendix}

\end{document}